# Engineering NV Centers via Hydrogen-Driven Defect Chemistry in CVD Diamonds for Quantum Applications: $NVH_x$ Dissociations into NV, Origin of 468nm Center, and Cause of Brown Coloration

Mubashir Mansoor[1,2*], Kamil Czelej[3], Sally Eaton-Magaña[4], Mehya Mansoor[1,2], Rümeysa Salci[5,6], Maryam Mansoor[2,7], Taryn Linzmeyer[4], Yahya Sorkhe[1], Kyaw S. Moe[8], Ömer Özyildirim[9], Kouki Kitajima[10], Mehmet Ali Sarsil[11], Taylan Erol[12], Gökay Hamamci[12], Onur Ergen[11], Adnan Kurt[12], Arya Andre Akhavan[13], Zuhal Er[14,15], Sergei Rubanov[16], Nikolai M. Kazuchits[17], Aisha Gokce[5], Nick Davies[18], Servet Timur[1], Steven Prawer[19*], Alexander Zaitsev[20*], Mustafa Ürgen[1*]

[1] *Metallurgical and materials engineering department, Istanbul Technical University, Istanbul, Türkiye*
[2] *STEMZY Mühendislik Çözümleri San. ve Tic. A.Ş., Istanbul Türkiye*
[3] *Faculty of Chemical and Process Engineering, Warsaw University of Technology, Waryńskiego 1, 00-645, Warsaw, Poland*
[4] *Gemological Institute of America, Carlsbad, CA, United States*
[5] *TUBITAK National Metrology Institute, Kocaeli, Türkiye*
[6] *Department of Physics, Gebze Technical University, Gebze, Türkiye*
[7] *Energy Institute, Istanbul Technical University, Istanbul, Türkiye*
[8] *Gemological Institute of America, New York, NY, United States*
[9] *Department of Computer Engineering, Istanbul Technical University, Istanbul, Türkiye*
[10] *Department of Geoscience, University of Wisconsin Madison, WI, United States*
[11] *Electronics and communications engineering, Istanbul Technical University, Istanbul, Türkiye*
[12] *Appsilon Enterprise BV, Delft, Netherlands*
[13] *Precision Gems Research, Toronto, Ontario, Canada*
[14] *Maritime Faculty, Istanbul Technical University, Istanbul, Türkiye*
[15] *Department of Applied Physics, Istanbul Technical University, Istanbul, Türkiye*
[16] *Ian Holmes Imaging Center, Bio21, University of Melbourne, Parkville, Victoria, Australia*
[17] *Department of Physics, Belarusian State University, Minsk, Belarus*
[18] *De Beers UK Ltd, Berkshire, United Kingdom*
[19] *School of Physics, University of Melbourne, Parkville, Victoria, Australia*
[20] *The College of Staten Island, City University of New York, New York, United States*

***Corresponding authors:**
Mubashir Mansoor: mansoor17@itu.edu.tr
Prof. Dr. Mustafa Ürgen: urgen@itu.edu.tr
Prof. Dr. Steven Prawer: s.prawer@unimelb.edu.au
Prof. Dr. Alexander Zaitsev: alexander.zaitsev@csi.cuny.edu

## Abstract

Achieving high NV center conversion efficiency remains a key challenge in advancing diamond-based quantum technologies. The generally accepted mechanism for NV formation is that irradiation-induced vacancies become mobile during annealing and are trapped by substitutional nitrogen. However, the suggested mechanism does not consider the presence and role of the presence of hydrogen in the diamond and its influence on the NV formation pathway. This is despite *ab initio* calculations which strongly suggest the formation of hydrogen-passivated NV centers during CVD diamond growth. Recent experimental observations showing a strong spatial correlation between NV centers, brown coloration, and the 468 nm luminescence center in as-grown CVD diamonds prompted us to investigate the atomistic origin of these phenomena in the presence of $N_xVH_y$-type complex defects. We used hybrid density functional theory (DFT) calculations and spectroscopic analysis of CVD diamonds grown with varying nitrogen content to investigate defect equilibria during growth. We identified the 468 nm center as the $NVH^-$ defect—a hydrogen-passivated NV center—and assigned the characteristic UV–VIS absorption bands at 270, 360, and 520 nm to $N_xVH_y$ complexes. Our findings reveal that hydrogen plays a central role in stabilizing these defects during growth. We further showed that $NVH_x$ complex defects dissociate into NV centers and interstitial hydrogen during post-growth irradiation and annealing, complementing vacancy trapping by substitutional nitrogen. These results provide a unified picture of the defect chemistry underlying brown coloration, 468 nm center and NV formation in CVD diamonds, offering new insights for optimizing diamond synthesis and processing for quantum applications by taking advantage of hydrogen's role and dissociation of $NVH_x$ complexes.

## 1. Introduction

Diamond has long been admired for its extreme physical properties such as high thermal conductivity, exceptional hardness, and wide bandgap. Recent advances in materials science have additionally positioned it as a key platform for emerging technologies [1,2]. From high-frequency power electronics to quantum photonic devices and single-photon emitters [3], the technological potential of diamond lies not only in its pristine lattice but in its ability to host a rich variety of optically and electronically active defects [4-6].

Among these, the nitrogen-vacancy (NV) center has emerged as a leading candidate for use in quantum information processing, capable of stable operation at room temperature. Its unique spin properties, long coherence times [7] and compatibility with optical readout protocols [8] have enabled a broad range of applications, including quantum sensing and quantum information processing.

A key obstacle to optimizing NV based diamonds for quantum and optoelectronic applications lies in the nature of their as-grown defect landscape due to conditions of the CVD growth, such as formation of secondary defects that negatively impact the performance of NV centers [9, 10], and the limited formation of NV from precursors such as N and V, which results in higher concentrations of N compared to NV. An extraordinarily high concentration of secondary defects and N tend to exhibit a brown coloration as well that is attributed to vacancy cluster defect formations [11]. This coloration is undesirable in multiple contexts: it reduces optical transparency, diminishes luminescence efficiency, and severely quenches the emission intensity of key color centers such as NV. Therefore, for applications in quantum optics, where high-fidelity single-photon emission is essential, these issues are critical and the brown coloration's atomistic cause may very well be hiding the tale of NV conversion efficiencies.

This brown coloration can be mitigated through post-growth treatments. High-temperature annealing above 1600 °C and electron irradiation followed by lower-temperature heat treatment in the 600–1000 °C range result in significant NV formation and enhanced coherence time. Therefore, electron irradiation followed by annealing at ambient pressure is the primary post processing method for NV formation in diamond [10, 12]. Upon electron irradiation vacancies are formed, which become mobile during annealing and can combine with substitutional nitrogen ($N_s$) to form NV centers, although NV centers can form through other means of irradiation as well [13]. Photoluminescence (PL) studies confirm a simultaneous rise in NV intensity and a visible shift in the color of the material – from brown to pink or brownish-pink – by following these procedures, which is associated with a rise in the UV-VIS absorption of 350 to 400 nm region after irradiation, followed by rise in 500 - 600 nm UV-VIS region [14,15]. Higher temperature annealing close to 2000 ℃ in turn can render the diamonds colorless [16, 17]. Therefore, the established correlations between transformations of brown coloration and the changing point defect landscape that triggers NV formation once again emphasizes the importance of understanding the atomistic cause of brown coloration and how it is linked to NV formation. The relationship may open up a fresh perspective on the NV formation pathways.

The underlying cause of the brown coloration in as-grown CVD diamond remains the subject of debate. Early and widely accepted hypotheses focus on extended defects, particularly spherical clusters of carbon vacancies, which were proposed to scatter light and produce the observed optical absorption [18-20]. Experimental support for this idea came from transmission electron microscopy (TEM) studies, which reported the presence of nanometer-scale clusters in select samples [21, 22]. However, the formation energies of such clusters are extraordinarily high [23] – on the order of 120 eV for 60-vacancy aggregates – and their spontaneous formation under equilibrium growth conditions is difficult to justify, and even if they do form under Ostwald ripening-type processes as proposed by Fujita et al [23], the subsequent annealing at high temperature is debatable given that the thermodynamic activity of vacancy increases at high temperature [24], also seen by rising concentration of multi-vacancy extended defects that undergo long HPHT treatments [25]. Moreover, the predicted electronic density of states and optical absorption spectra from *ab initio* models of these clusters do not match experimental data, particularly the characteristic UV–VIS absorption bands at 270, 350, and 520 nm [26]. While Mie scattering could explain the broad featureless absorption background seen in some spectra, it fails to account for these discrete broad bands.

A spatial correlation exists between brown coloration and NV formation, which has long been viewed as coincidental – an artifact of overlapping processing steps [27]. However, growing evidence suggests that these two phenomena may not be independent, but rather manifestations of a shared defect evolution pathway. Zaitsev et al. [27] has found a direct correlation between the nitrogen content in the CVD chamber with the intensity of brown coloration. Moreover, an intriguing development in this area is the identification of a 468 nm photoluminescence center, which is directly correlated to the intensity of the brown coloration in as-grown CVD diamond. First reported by Zaitsev et al., [27] its significance and correlation with the brown color is shown to be such that the 468 nm center has been termed "the luminescence of brown CVD diamond". However, the origin of this center is not yet definitively known and remains ambiguous. Based on its phonon sideband, a vacancy-related origin has been proposed and based on its relationship to brown coloration and spatial proximity to NV centers in as-grown material, suggests that the these optical phenomena could be linked to a shared family of defects, especially given that the 468 nm center and the brown coloration diminish in intensity upon annealing in a correlated manner, and prior study by Rzepka et al. [28] shed doubt on its nitrogen related nature.

In other words, the defects responsible for the brown color may either compete with or evolve into NV centers during post-treatment. Understanding this relationship is therefore essential, not only to improve NV conversion efficiency, but also to enable more reliable control over the optical and electronic properties of diamond-based devices. Therefore, understanding the defect equilibria during growth and subsequent treatments is of utmost importance.

Nitrogen and hydrogen are ubiquitous in CVD growth environments – hydrogen as the primary process gas (typically >90%), and nitrogen is an intentional additive to increase growth rates, modify morphology and a necessary ingredient for NV formation. The incorporation of H and N into the diamond lattice is therefore commonly seen. SIMS analysis [29] on polycrystalline CVD diamond has revealed H-concentrations as high as $3\times10^{19}$ $cm^{-3}$ and confirmed significant H trapping within the diamond grains, therefore solubility of H in single crystal diamond is expected and the H-related color centers are ubiquitous [24]. Attempts are also made to document nitrogen and hydrogen related point defect complexes and their potential spectroscopic signatures [30, 31]. Therefore, considering studies that have demonstrated a direct correlation between nitrogen content in the gas phase and the intensity of the brown coloration [16, 32,33], and impact of UV exposure and mild thermal treatments that modulate the charge states of substitutional nitrogen ($N_s^0$ and $N_s^+$), producing corresponding shifts in the absorption spectra and brown coloration [34], may suggest that nitrogen, possibly in complex with hydrogen and vacancies, plays a central role in defining the optical signature of brown CVD diamond. The 520 nm UV-VIS band was indeed assigned to the NVH defect by Khan et al [34] and the research by Meng et al. [16] proposes the 350-400 nm band and the increasing continuum towards the shorter wavelengths may be hydrogen related as well. There is therefore an ongoing trend to assign unknown spectral peaks of various nitrogen and hydrogen defects to their particular atomistic cause [35].

Taken all together, these findings point to a broader, interconnected defect chemistry involving nitrogen, hydrogen, and vacancies, the brown coloration, the 468 nm luminescence, and the evolution of NV centers under post-treatment, all of which may reflect different states or transformations within the same defect family. Identifying the specific atomic configurations responsible is therefore more than a matter of academic interest – it has direct implications for tuning NV conversion efficiency, optimizing diamond for quantum applications, and controlling the optical properties of engineered crystals.

In this work, we address this challenge through a combined theoretical and experimental approach. Using first-principles calculations, we identify nitrogen–hydrogen–vacancy complexes that are thermodynamically and kinetically plausible under CVD growth conditions and elucidate on their annealing mechanism. These predictions are tested against a series of experimental studies, including photoluminescence spectroscopy, secondary ion mass spectroscopy, transmission electron microscopy, UV–VIS absorption, strain measurements, and post-treatment evolution of the defect equilibria. Our results shed new light on the atomistic origin of the 468 nm center, clarify its connection to brown coloration, and provide insight into the pathways for efficient NV center formation. Together, these findings contribute to a more complete understanding of defect engineering in diamond and pave the way for improved materials design in quantum technologies.

## 2. Results & Discussions

### *2.1. Defect equilibria during diamond growth and their consequences*

CVD diamond growth typically occurs under a hydrogen-rich atmosphere, with nitrogen widely used as a catalyst to enhance the growth rate [36]. The primary precursors $H_2$ and $CH_4$ dissociate into atomic hydrogen and methyl radicals via microwave plasma or alternative activation methods. As a result, both atomic hydrogen and nitrogen are commonly incorporated into the growing diamond, unless grown under extremely controlled conditions that contain implantation of N in pristine diamond. Therefore, for most practical purposes, the role of N and H in defect chemistry is critical. While the effects of nitrogen have been extensively investigated [37], the complex interplay of nitrogen–hydrogen defects remains underexplored, despite signs regarding their possible contributions in defect landscape of CVD diamonds [24, 38-40]. The role of hydrogen is particularly important to consider due the generally high mobility of interstitial hydrogen in comparison to substitutional covalently bonded atoms or vacancies, as seen in diamond and other materials [40, 41, 45]. Therefore, given enough time and energy, interactions of H and N are inevitable. Our *ab initio* thermodynamic calculations reveal the central role of atomic hydrogen in shaping macroscopic properties and defect evolution, particularly in processes related to NV center formation, brown coloration, and the emergence of elusive color centers such as those emitting at 468 and 596 nm.

Hydrogen promotes the formation of $NVH_x$ and trapping of H by $N_2V$ (H3 centers) is also plausible, to form $N_2VH_x$ defects, which exhibit significantly lower formation energies than the NV centers (Fig. 1a). For example, $NVH_2$ and $NVH_3$ are energetically favorable as compared to isolated substitutional nitrogen (C-center), indicating that hydrogen-passivated NV and $N_2V$ centers can form in substantial concentrations during growth (Fig. 1a). In these complexes, the vacancy serves as a hydrogen trap; hydrogen passivation of carbon dangling bonds lowers the Gibbs free energy, increasing thermodynamic stability and concentration of the hydrogenated defect structures during growth. Amongst these, $NVH_3$ is the most stable, having the lowest formation energy in the $NVH_x$ group, with concentrations far exceeding the single substitutional N (C-center) and NVH. However, given its complete electronic passivation, this center will not be detectable through photoluminescence or UV-VIS spectroscopy. We also find excessive nitrogen concentration ($\mu_N^{N_2} - 1eV < \mu_N \ll \mu_N^{N_2}$)in presence of hydrogen plasma will render the diamond phase unstable and phase decomposition commences at temperatures exceeding 1300 K (see section 1 of supplementary file).

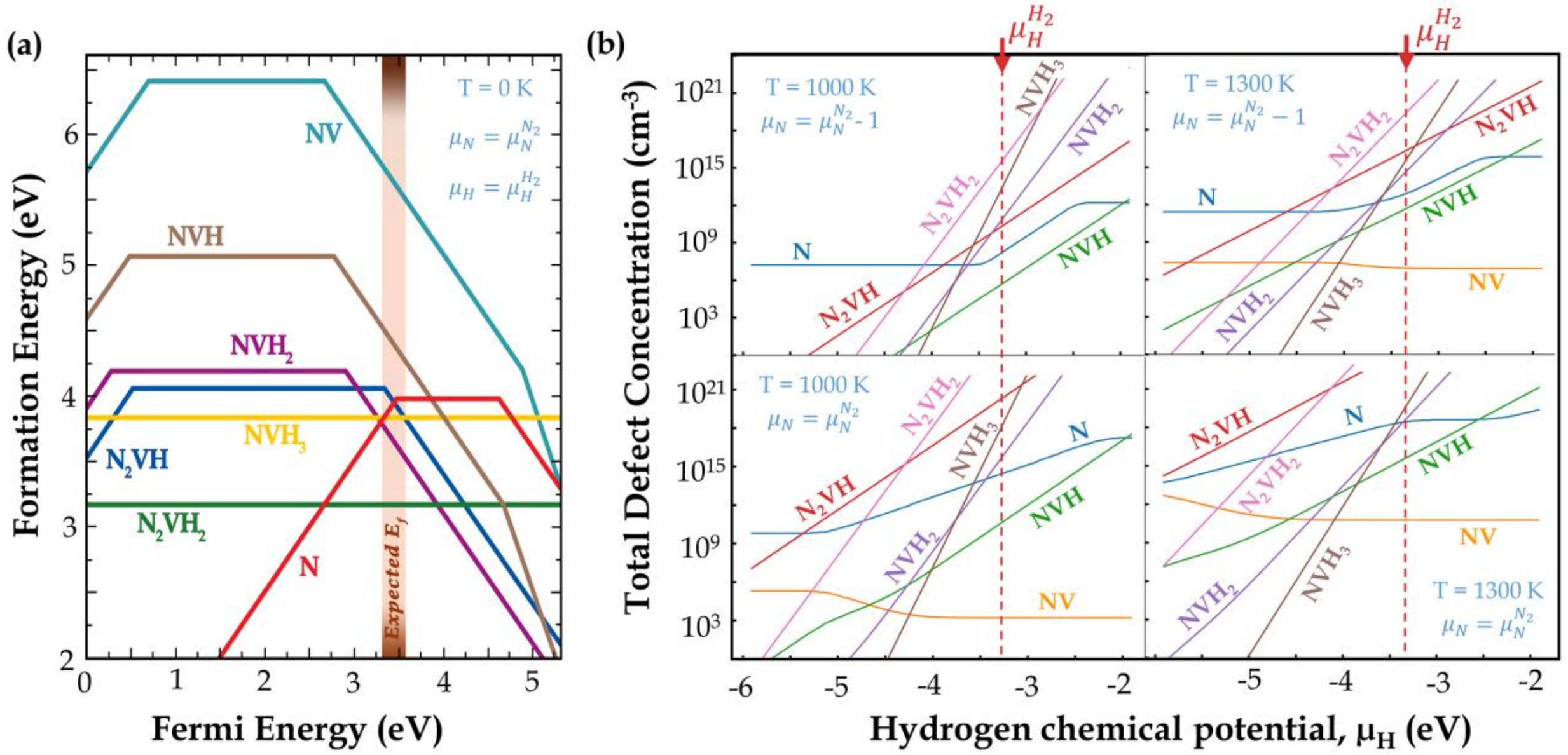

***Fig. 1.*** *(a) The formation energy of hydrogen passivated NV and $N_2V$ (H3) centers are calculated as function of equilibrium Fermi energy, which shows their significance during diamond growth. The expected equilibrium Fermi is estimated at 3.2 to 3.6 eV for achieving charge neutrality, signifying the dominance of fully hydrogen passivated centers. (b) The equilibrium concentrations calculated as a function of temperature and varying dopant chemical potentials show the dominance of $N_2VH_2$, $NVH_3$ and $NVH_2$ over N (C-center) during diamond growth. The dashed red line shows the defect equilibria referenced to hydrogen molecule's chemical potential, however, the hydrogen chemical potential in the system depends on plasma conditions. Phase decomposition of diamond is also predicted at temperatures exceeding 1300 K if the thermodynamic activity of nitrogen is too high (i.e. if the chemical potential exceeds $\mu_N^{N_2} - 1eV$).*

Theoretical modeling also demonstrates that hydrogen passivation of NV centers induces considerable tensile stress, with local atomic strains on the order of $10^{-2}$ (1%). As more hydrogen atoms occupy the vacancy site, the induced strain increases – an expected consequence of electrostatic repulsion between hydrogen atoms, which expands the vacancy volume. These local hydrogen-induced tensile stresses are substantial: ~10 GPa for NVH and up to ~45 GPa for $NVH_3$ (Fig. 2). The hydrogen-induced strain in this context is quite similar to the strain H causes in other materials [41]. The H-induced tensile strain is also well-known in the CVD nano-diamonds, however, the cause of which was previously only attributed to grain boundary effects [42]. In light of these results, we also underline the potential impacts of nitrogen-hydrogen related complex point defects on not just the strains in the diamond structure but also on the optoelectronic behavior of diamond. Our Raman spectroscopy measurements on as-grown diamond sample with two regions of lighter and darker brown, from surface to depths of approximately 300 µm, show that the first order diamond Raman line deviates by -0.6 $cm^{-1}$ between dark brown and near-colorless regions (Fig. 2g), corresponding to an average tensile stress of ~1.7 GPa across the lattice (see Section 2 of Supplementary Data). This  correlation between potentials of thermodynamically stable hydrogen related defects for causing tensile strain, and the experimentally seen higher tensile strains in darker brown regions suggests a possible hydrogen related nature for the brown coloration.

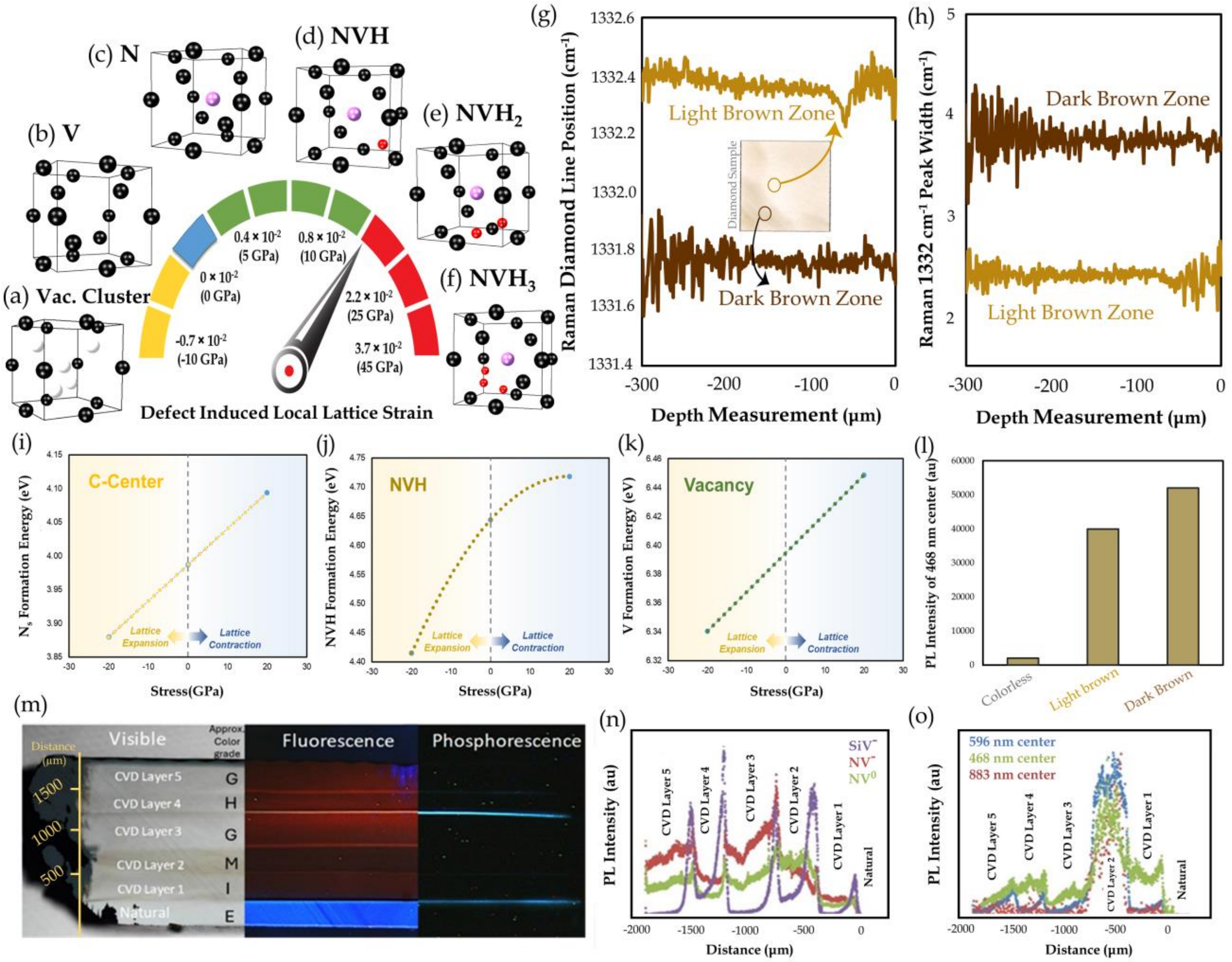

***Fig. 2.*** *The impact of local atomic strains are calculated which are; (a) Vacancy clusters causes compression of varying magnitidues depending on the size of the cluster, (b) Single carbon -vacancy shows compression under extended vacancy defects, however, single vacancies contribute negligibly to strain, (c) N is tensile at approximately* $0.4\times10^{-2}$*, (d) NVH is tensile with nearly 5GPa of local stress, (e)* $NVH_2$ *is tensile with nearly 25 GPa of atomic scale strain and (f)* $NVH_3$ *shows the maximum tensile strain of* $3.7\times10^{-2}$ *corresponding to 45 GPa. (g) Diamonds of higher intensity brown coloration experimentally show a tensile strain in the bulk of the crystal, which exceeds 1.7 GPa, as measured by Raman spectroscopy through the deviations in the position of the Diamond-band. (h) The tensile strain is also associated with peak broadening in the darker brown regions of grown diamond crystal. The tensile strain is also associated with theoretically calculated changes in formation energies of (i) N, (j) NVH and (k) Vacancy centers, meaning their expected higher concentrations due to hydrogen induced lattice expansion. (l) The intensity of brown coloration is also associated with the intensity of the 468 nm center. (m) Optical image of the diamond crystal grown with stratified nitrogen concentration corresponding to varying gemological color grades from E to M under visible light, and UV light is shown. (n) The diamond demonstrates peaks in the intensity of formed* $NV^0$*,* $NV^-$ *and* $SiV^-$*, which correspond to peaks in as-grown (o) 596, 468 and 883 nm centers.*

The emergence of measurable macroscopic tensile strain from point defects present at only ppm concentrations highlights the profound influence of sub-nanometer-scale stress fields – fully consistent with our *ab initio* predictions. Crucially, this tensile behavior stands in direct contradiction to models invoking extended vacancy clusters, which are expected to induce lattice compression, not expansion (see supplementary data Fig. S5 and Fig. 2a) [43, 44]. These findings cast significant doubt on the significance of vacancy-cluster hypothesis for brown coloration, and the emergence of tensile strain in darker brown regions of as-grown CVD diamonds instead point toward possible contributions from nitrogen and hydrogen related point-defect-driven origins.

The hydrogen-induced tensile strain has key consequences: it increases nitrogen solubility and reduces the formation energy of key defects (e.g., C-center, NVH, vacancies, see Fig. 2i-2k). For NVH a local tensile strain of 1% equates to an estimated 0.25 eV reduction in the defect formation energy, which equals to an increased concentration of $e^{\left(\frac{0.25}{k_B T}\right)}$ that is approximately one order of magnitude increase in NVH under a growth temperature 1000 °C. Thus, the strain generated by nitrogen–hydrogen complexes contributes to higher overall defect densities, which corresponds to regions of enhanced brown coloration, and increased 468 nm luminescence as seen experimentally. Experimental studies on CVD-grown single crystals with varying nitrogen content confirm these trends: Raman spectra show Diamond-band downshifts and FWHM broadening (Fig. 2g and 2h), both indicators of strain, and correlate strongly with increases in brown color intensity and 468 nm PL emission. In fact, our data appears to indicate that 468 nm luminescence intensity is a reliable proxy for hydrogen-induced tensile strain in CVD diamond (Fig. 2g and 2l).

In diamonds grown with stratified nitrogen concentrations, we observe a stepwise increase in the concentration of $NV^0$, $NV^-$, SiV, 596 nm and 883 nm defects, tightly following the PL intensity of the 468 nm center (Fig. 2n and 2o). This provides direct experimental validation of our theoretical predictions: hydrogen-induced lattice strain lowers defect formation energies (Fig. 2i – 2k), promotes their incorporation, and amplifies the optical features associated with brown coloration. These insights offer a cohesive framework linking local tensile stress, nitrogen-hydrogen related defect chemistry, and 468nm PL and the brown coloration in CVD diamond.

### *2.2.Atomistic origin of the 468 nm center*

The observed correlations between strain, the 468 nm luminescence, and the brown coloration in as-grown CVD diamonds – along with the spatial proximity of NV centers to this elusive PL center – suggest that uncovering the origin of the 468 nm emission could significantly advance our understanding of the defect transformation pathways underlying NV formation and brown coloration.

The binding of hydrogen to NV is exothermic across most of the electron chemical potential range ($E_x$ > 1.5 eV), making NVH a likely product during growth at approximately 1000 °C. In other words, the vacancy of the NV center acts as a hydrogen trap during diamond growth and formation of NV centers with varying degrees of hydrogen passivation are inevitable (NVH, $NVH_2$ and $NVH_3$). These theoretical results are consistent with experimental reports of NV passivation under hydrogen plasma, as previously demonstrated by Stacey et al. [45], possibly indicating the conversion of NV into $NVH_x$ type defects.

Given the equilibrium Fermi level in intrinsic CVD diamond – typically between 3.2 and 3.6 eV, due to near-equal concentrations of $N^0$ and $N^+$ within one order of magnitude [46]– both NV and NVH are expected to coexist during growth in hydrogen-plasma environments. This is a direct outcome of preserving charge neutrality (see supplementary file for more details). Therefore, considering the thermodynamic charge transitions of NVH calculated through highly accurate hybrid density functional (HSE06) and the experimental observations by Zaitsev et al. [27] on the spatial proximity of NV and the 468 nm luminescence pave the way for a hypothesis on the NVH related origin of this luminescent center, which we investigate in detail as follows.

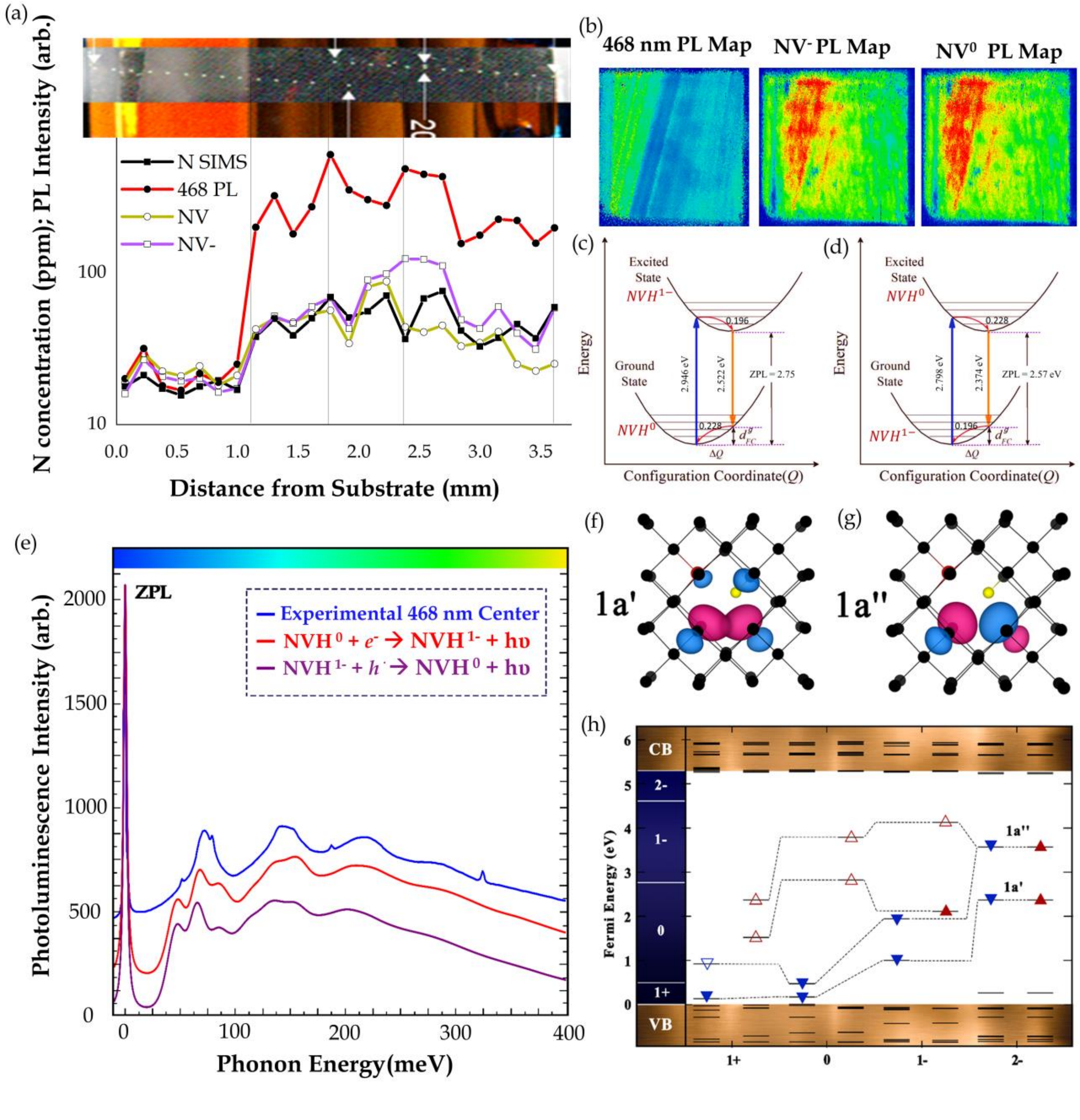

***Fig. 3.*** *(a) The secondary ion mass spectroscopy (SIMS) results coupled with photoluminescence data from the same regions of diamond shows a direct correlation between the intensity of brown coloration, 468 nm PL and the nitrogen content of the diamond. (b) The special proximity of the 468 nm luminescence with the NV*$^0$ *and NV*$^-$ *are also seen through photoluminescence mapping of an as-grown diamond. The configuration coordinate diagrams of the NVH defect for the (c) electron-capture and (d) hole-capture reveal ZPL lines that are within the typical error of DFT for explaining the 468 nm center. (e) The PL lineshape of the 468 nm center as observed experimentally by Zaitsev et al [27], and NVH Pl lineshape we calculated by DFT are compared, which exhibit near exact matching of phonon sideband. (f) The symmetric 1a' and (g) asymmetric 1a'' defect-related orbitals are shown, along with the (h) Kohn-Sham levels of the NVH defect for different charge states.*

The NVH can be stable in 2-, 1-, 0 and 1+ charge states, with two localized defect levels are present in the band gap: the symmetric 1a' and asymmetric 1a". The 3D plot of 1a' and 1a" orbitals (Fig. 3f, 3g) indicates that the former is localized on a carbon dangling bonds and partially on nitrogen and C-H bond, whereas the latter is a pure C dangling bond state. In the positive charge state, the 1a' is occupied by one electron and the 1a" leading to a paramagnetic doublet S=1/2 spin state, but due to its narrow stability window the formation of $NVH^{1+}$ is highly unlikely. The neutral NVH exhibits S=1 triplet 3A" many body electronic state with half occupied 1a' and 1a" levels in spin down channel. Trapping of an additional electron on the 1a" level leads to the $NVH^{1-}$, which is a spin doublet. This charge state exhibits possible spin conserving internal transition 2A" → 2A' in the spin minority channel that falls within IR range. We computed the PL lineshape for this transition to mediate its future identification (see supplementary files). As can be seen, the simulated phonon sideband associated with 2A" → 2A' transition and ZPL value differ from the experimentally measured PL sideband of 468 nm ZPL emission. Finally, the doubly negative charge state of NVH is a closed shell singlet with both 1a' and 1a" states fully occupied. Prior investigation by Deak et al. [47] had underestimated the adiabatic charge transition of $NVH^{0/-}$ to be 2.4 eV, most likely due to the empirical charge correction scheme that they used, which is known to be prone to errors. Such errors can be avoided only through fully *ab initio* charge correction methods that were developed much later than Deak’s attempt to calculate the transitions of NVH. Moreover, the experimental study of Khan et al [34] on the NVH makes it clear that the exact value for this transition is not well defined either.

To further probe the nature of the 468 nm center, we conducted secondary ion mass spectrometry (SIMS) and found a direct correlation between the 468 nm center’s PL intensity and doped nitrogen content, similar to the behavior of the NV center, and reinforcing the nitrogen-related origin of the defect. Moreover, the enhanced nitrogen doping causes a rise in equilibrium Fermi energy, which promotes evolution of negative charge states in most $N_xVH_y$ defect complexes (Fig.1a). We therefore carried out a computational analysis on the PL lineshape of the NVH based on the methodology of Alkauskas et al. [48]. The resulting theoretical photoluminescence spectral profile of the NVH including the phonon sidebands and quasi-local vibrational modes – closely resembles those reported experimentally by Zaitsev et al. [27] for the 468 nm center (Fig. 4a). The nature of 468 nm optical line is associated with bound exciton transitions where the hole localized at the nondegenerate VBM state radiatively recombines with electron occupying 1a'' state in spin minority channel of $NVH^{-}$, or the electron localized in CBM state recombines with hole occupying the 1a'' state in spin minority channel of $NVH^{0}$. The phonon replica for both transitions are nearly identical because these transitions involve the same localized orbital in the bandgap, leading to similar structural relaxation upon optical transition. In general, bound exciton transitions do not produce sharp and clearly distinguishable ZPL but is some cases where the Huang-Rhys factor is small enough it can be observed. Similar bound exciton transition with clearly visible ZPL followed by distinct phonon sideband has been recently reported for C center in silicon [49].

To independently verify the NVH assignment, we compared the spatial distributions of 468 nm PL and the 3123 $cm^{-1}$ FTIR peak, a known vibrational signature of the $NVH^{0}$ defect that is consistently reported in brown and brownish pink colored CVD diamonds [16, 50, 51]. We find that in CVD diamonds grown with varying nitrogen content, both signals co-varied: higher 468 nm intensity coincided with stronger 3123 $cm^{-1}$ absorption, especially in regions exhibiting brown coloration (Fig. 4a). In contrast, colorless zones reveal a marked reduction in both features, underscoring their shared atomistic origin.

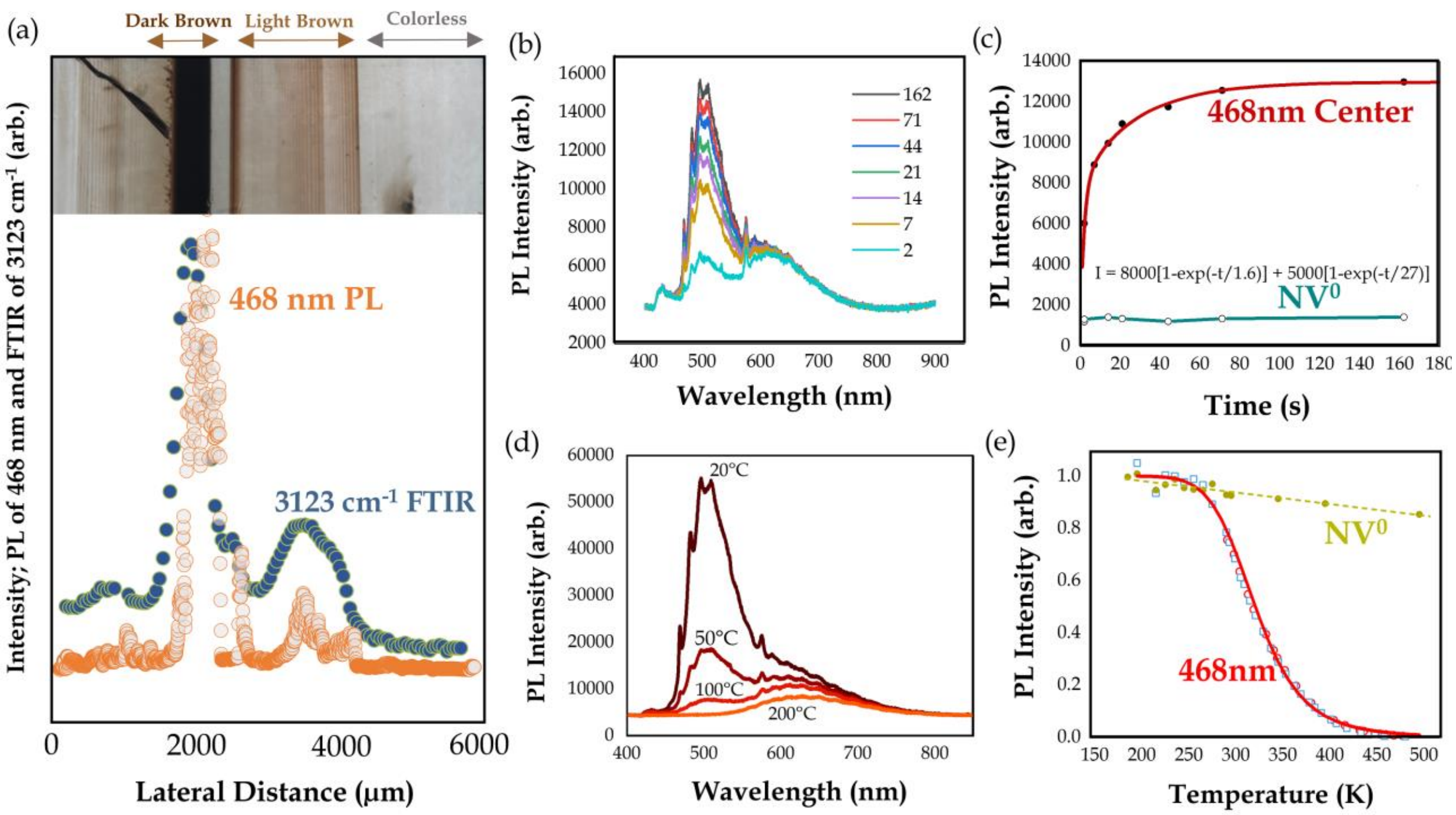

***Fig. 4.*** *(a) A line-scan of the 468 nm PL signal and the well-known 3123cm*$^{-1}$ *FTIR signature of NVH are shown on a stratified diamond crystal with varying intensities of brown coloration and nitrogen content. Both signals show a direct correlation. (b) The increase in 468 nm PL intensity under UV illumination for 2 to 162 seconds is shown, which shows an (c) asymptotical behavior with UV exposure time. The same behavior under UV is not seen for NV*$^0$. *(d) The substantial reduction in 468 nm PL intensity is seen upon heating of diamond, however, (e) the impact of temperature is comparatively insignificant for NV*$^0$.

Given the near-identical PL lineshape between the experimentally observed 468 nm center and our calculated NVH spectrum, the consistency of the ZPL energy with DFT predictions, the nitrogen-related nature of the defect from SIMS analysis, and the co-localization of the 468 nm emission with the well-established FTIR signature of NVH (3123 cm$^{-1}$), we suggest the 468 nm luminescence is a bound exciton transition of the NVH defect.

To determine the active charge state, we studied the response of the 468 nm center to external stimuli. Upon UV illumination, the 468 nm PL intensity increases, while thermal annealing reduces it (Fig 4b-4e) – consistent with the behavior imposed by a negatively charged defect. UV exposure raises the Fermi level, favoring the formation of NVH$^-$, while annealing lowers the Fermi level and diminishes this state. This also explains why the NV$^0$ is not affected by the UV exposure; its formation energy is independent of the Fermi energy (Fig. 1a). These trends confirm that the 468 nm emission arises from the $NVH^{-}$ charge state and corresponds to bound exciton transition.

### *2.3. The nitrogen and hydrogen induced brown color*

The previously suggested link between the 468 nm luminescence and brown coloration – attributed to vacancy clusters – has been re-evaluated in light of our findings that assign the 468 nm center to the $NVH^{-}$ defect and the observations of tensile strains in brown-rich regions that demonstrate the significance of hydrogenated defects rather vacancy clusters which should cause lattice compressions. We therefore calculated the dielectric functions of the $N_xVH_y$ defect complexes and incorporated their equilibrium concentrations under CVD growth conditions in a partition function to simulate their cumulative contribution to the UV–VIS absorption spectrum.

This approach successfully reproduces both the overall brown coloration and the characteristic broad absorption bands observed experimentally at ~270, 350, and 520 nm (Fig. 5a and 5e-5g). Our results reveal that these bands originate not from a single defect, but from overlapping electronic transitions across multiple $N_xVH_y$ complexes. Specifically, the 270 nm band previously attributed to $N^{0/-}$ transition (electronic transition between the valence band and the N donor level), and the 520 nm band linked to NVH are confirmed to involve contributions from several defect species, as summarized in Table 1. The superposition of these transitions in accordance with their equilibrium concentrations explains both the spectral width and the variability of absorption profiles in different samples.

***Table 1.*** *The defects contributing to the transitions associated with the 270, 360 and 520 nm absorption bands of as-grown CVD diamond*

| **Absorption Band (nm)** | **Responsible Defects** | **Adiabatic transition (eV)** |
|---|---|---|
| 270 | $N^{(0/-)}$ | 4.58 |
| | $NVH^{(0/+)}$ | 4.85 |
| | $NVH_2^{(0/+)}$ | 5.05 |
| | $NVH^{(-/2-)}$ | 4.60 |
| 360 | $N^{(+/0)}$ | 3.44 |
| | $NVH_2^{(0/-)}$ | 2.88 |
| | $N_2VH^{(0/-)}$ | 3.30 |
| 520 | $NV^{(0/-)}$ | 2.64 |
| | $NV^{(-/0)}$ | 2.68 |
| | $NVH^{(0/-)}$ | 2.74 |
| | $NVH^{(-/0)}$ | 2.58 |
| | $NVH_2^{(-/0)}$ | 2.44 |

Previous work by Khan et al. [34] demonstrated that the brown coloration in as-grown CVD diamond exhibits pronounced thermochromic behavior; under UV illumination, the coloration deepens, whereas subsequent heating results in a marked lightening of the material and a corresponding decrease in absorption at ~270, 360, and 520 nm (Fig. 5). These observations, though previously unexplained at the atomistic scale, are now directly accounted for by our assignment of the brown coloration to charge transitions of nitrogen–vacancy-hydrogen ($N_xVH_y$) point defect complexes. Our theoretical framework shows that UV exposure increases the equilibrium Fermi level, thereby lowering the formation energy of negatively charged $N_xVH_y$ defects and favoring their population. As the concentration of these optically active defects rises, so does the intensity of the brown coloration and the associated UV–VIS absorption bands. Conversely, thermal treatment induces a reduction in the equilibrium Fermi energy, increasing the formation energy of these negatively charged states and driving their partial deactivation or dissociation. This thermally induced reduction in defect population is reflected in the diminished absorption coefficient across the 270–500 nm spectral region (Fig. 5a), consistent with experimental spectra (Fig 5c). This is analogous to the experimentally seen effect on the intensity of the 468 nm center under UV and heating stimuli mentioned earlier.

Given that the fully *ab initio* approach employed in this study accurately reproduces the experimentally observed UV–VIS absorption spectrum and explains the previously ambiguous phenomena, we sought experimental verification on any possible changes in vacancy clusters that are considered in the literature as the cause of the brown coloration. Therefore, we performed post-growth thermal treatments and microstructural analysis on CVD diamonds exhibiting strong brown coloration and subjected them to low-pressure high-temperature (LPHT) annealing, resulting in diamonds of lower color intensity to colorless material. To evaluate whether this color transformation involved the dissolution or reconfiguration of extended defects such as vacancy clusters, we performed high-resolution transmission electron microscopy (HR-TEM) and electron energy loss spectroscopy (EELS) on pre- and post-annealed samples.

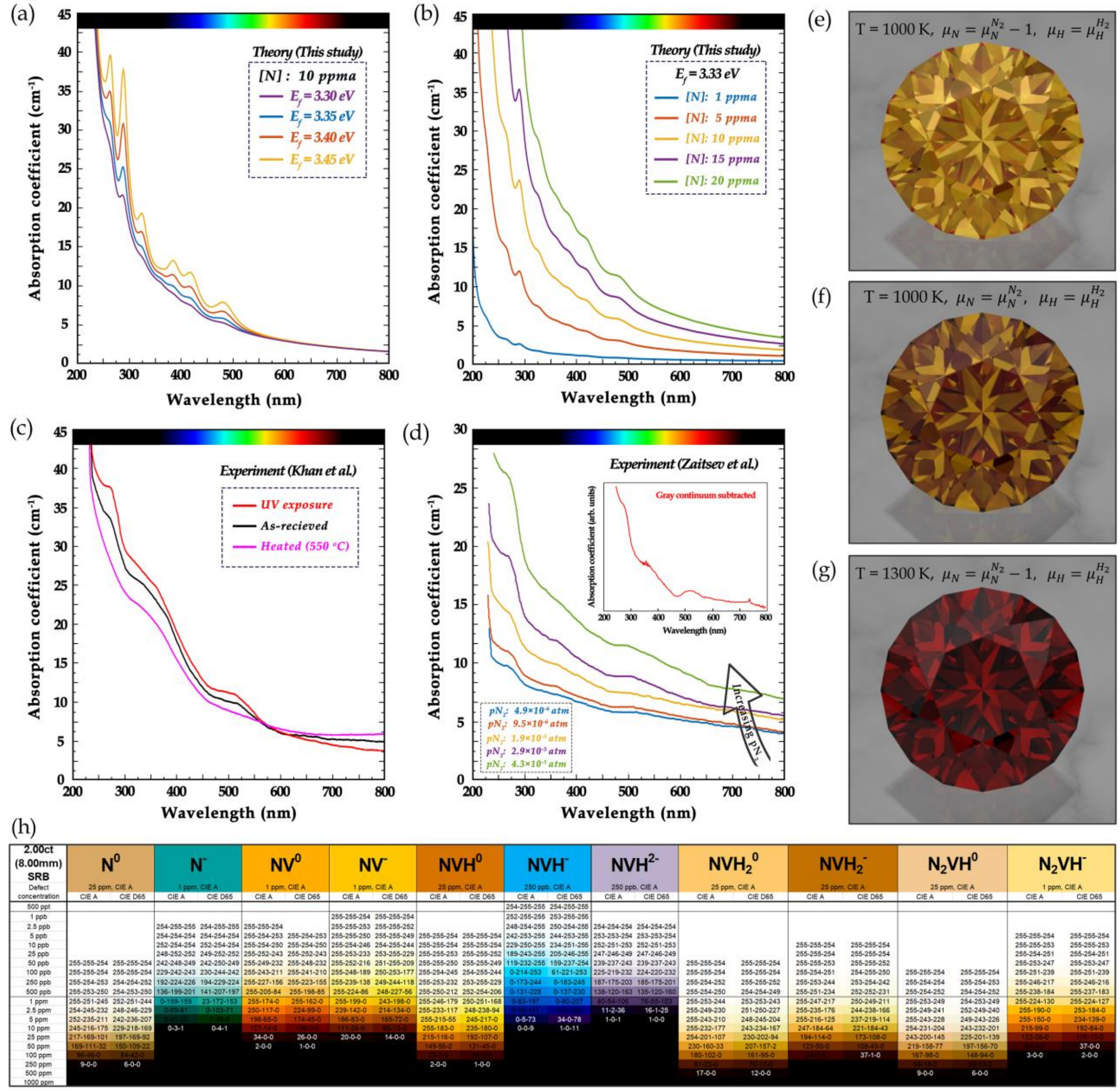

***Fig. 5.*** *(a) Equilibrium defect concentrations change as a function of equilibrium Fermi energy, and the impact on UV-VIS absorption as calculated by hybrid DFT is shown. Increasing Fermi energies causes a rise in the 270, 360 and 520 nm bands. (b)The theoretical UV-VIS spectra based on nitrogen and hydrogen defect complexes for varying nitrogen concentrations at equilibrium Fermi energy of 3.33 eV can reproduce the experimentally seen absorption in as-grown CVD diamonds. (c) the UV-exposure's increasing 270, 350 and 500 nm bands reported by Khan et al. and (d) experimentally grown diamonds of Zaitsev et al. under increasing nitrogen partial pressure. Images of diamonds generated from renders of calculated RGB based on hybrid-DFT calculations of the dielectric function are shown, which mimic the as-grown nitrogen doped CVD diamonds for (e) 1000K and* $\mu_N^{N_2} - 1$, *(f) 1000K and* $\mu_N^{N_2}$, *(g) 1300* $\mu_N^{N_2} - 1$.*(h) The RGB for each individual defect is calculated for a 2ct round brilliant cut diamond as a function of the defect concentrations, showing NVH⁻ primarily induces a blue coloration and the color of NV⁻ without the contribution of its reddish luminescence, is yellow.*

These investigations revealed no discernible structural features attributable to vacancy clusters, neither before nor after annealing. The only notable microstructural change was the gradual annealing of extended dislocations with increasing temperature, as detailed in the Section 2 of Supplementary Information and Fig. S6. Therefore, in light of these combined theoretical and experimental results, we attribute the brown coloration of as-grown CVD diamond primarily to optically active nitrogen-vacancy-hydrogen point defect complexes.

The presence of vacancy clusters, characterized by increased positron annihilation lifetime (PAL) of nearly 400 ps in as-grown brown diamonds (compared to ~110 ps in colorless diamonds) remains to be explained, as the PAL is reduced in proportion to the loss of brown color intensity during annealing [52, 53]. Although positron annihilation spectroscopy is a powerful tool for detection of vacancy clusters in material, it detects such vacancy clusters only indirectly, via the reduced local electron density that traps positrons and prolongs their lifetimes. Therefore, any mechanism that produces local tensile strain – and hence lowers electron density – can have a similar effect. Apart from vacancy clusters, hydrogen induced tensile strains lower electron density in a similar fashion as well and tend to prolong the PAL significantly (see supplementary data). Therefore, the long PAL in as grown brown CVD diamonds may be suggesting the tensile strains induced by the hydrogenated defects, which can explain the 468 nm luminescence and the brown coloration bands as well as the tensile strains seen by Raman mapping. One striking example of hydrogen-induced strain prolonging the positron lifetime is found in high-strength steel: hydrogen precharging of tempered martensitic steel under tensile deformation increased the positron lifetime from ~304 ps to up to ~400 ps [54], similar to the observations of Sakaki et al. [55]. In both cases, hydrogen induces significant tensile strains as seen in other steel phases too [41, 56], demonstrating that hydrogen-induced lattice dilation can indeed produce PAL value increase that is comparable to those seen between colorless and brown diamond. Regardless of the vacancy cluster or hydrogen-induced tensile strain related origins of the increased positron annihilation lifetime in brown as-grown CVD diamonds, the ab-initio modeling of the UV-VIS absorption spectra shown in Fig. 5a and the RGB color of individual defects show the dominance of the nitrogen-hydrogen related defects for reproducing the experimentally seen brown color and how a variety of N-V-H defect ensembles can cause brown coloration.

### *2.4. NV formation through dissociation of NVHx defects & annealing of brown color*

As demonstrated in Section 2.1, the formation of hydrogen-passivated NV centers ($NVH_x$) during diamond growth is thermodynamically favored under typical CVD conditions. Among these, NVH, $NVH_2$, and $NVH_3$ emerge as particularly relevant due to their exceptionally low formation energies – especially $NVH_3$, which is predicted to occur in concentrations exceeding that of isolated substitutional nitrogen (C-centers). The prevalence of $NVH_x$ in as-grown CVD diamond poses a direct challenge to the prevailing NV center formation model, which assumes NV centers mainly arise when vacancies and substitutional nitrogen migrate and pair during post-growth annealing (N + V → NV) [57, 58]. In the traditional model, one creates NV centers by first generating vacancies (e.g. via irradiation) and then thermally mobilizing those vacancies to pair with stationary N atoms. However, as shown thermodynamically in section 2.1, CVD growth itself incorporates nitrogen and hydrogen simultaneously, and the N-V-H defect complexes give rise to the 468 nm PL, tensile strains and the absorption bands of the brown coloration, and most importantly N–V pairs that already exist during growth, but are simply rendered optically and magnetically inactive due to hydrogen passivation. This means the standard post-growth vacancy–nitrogen pairing mechanism overlooks a reservoir of "pre-formed" NV centers that are hidden as $NVH_x$ complexes. Recent work has highlighted that NVH formation competes with NV formation in hydrogen-rich CVD processes, thereby significantly lowering the N-to-NV conversion efficiency [59]. For example, Findler et al. [59] show that NVH complexes outnumber NV centers by at least an order of magnitude in plasma CVD diamond, directly implying that many nitrogen defects that could form NVs instead become $NVH_x$ centers. This hydrogen passivation effect was also demonstrated by Stacey et al. [45] who observed that introducing hydrogen to NV-rich diamond caused a marked depletion of NV photoluminescence, consistent with the conversion of NV centers into non-radiative NVH complexes. In essence, hydrogen incorporation during growth sequesters NV centers as $NVH_x$, meaning the absence of NV luminescence in as-grown CVD diamonds does not necessarily indicate a lack of N–V pairs – rather, those pairs exist in a hydrogenated state. This insight complicates the conventional narrative and suggests that post-growth treatment is needed not just to form NVs, but to activate them by releasing hydrogen, which complements and occurs in parallel to the conventionally accepted mechanism of N-V pairing.

To investigate the fate of $NVH_x$ defects during post-growth processing, we performed *ab initio* DFT-based transition-state searches using the nudged elastic band (NEB) method, along with finite-temperature molecular dynamics simulations. Our results reveal a distinct mechanism for NV formation; the dissociation of $NVH_x$ defects into NV centers and mobile interstitial hydrogen (Fig 6). Upon gradual energy input via thermal or irradiation stimuli, hydrogen atoms gain sufficient kinetic energy to overcome the potential barrier posed by the NV trap in potentially as short as 75 femtoseconds. This process leaves behind a stable NV center, while the liberated hydrogen becomes interstitial or migrates to nearby vacancy sites (see Supplementary Video). The results provide a theoretical framework that is in direct agreement with the prior experimental work of Chakravarthi et al. [60], which proposed $NVH_x$ dissociation into NV through confocal tracking of NV formations during annealing.

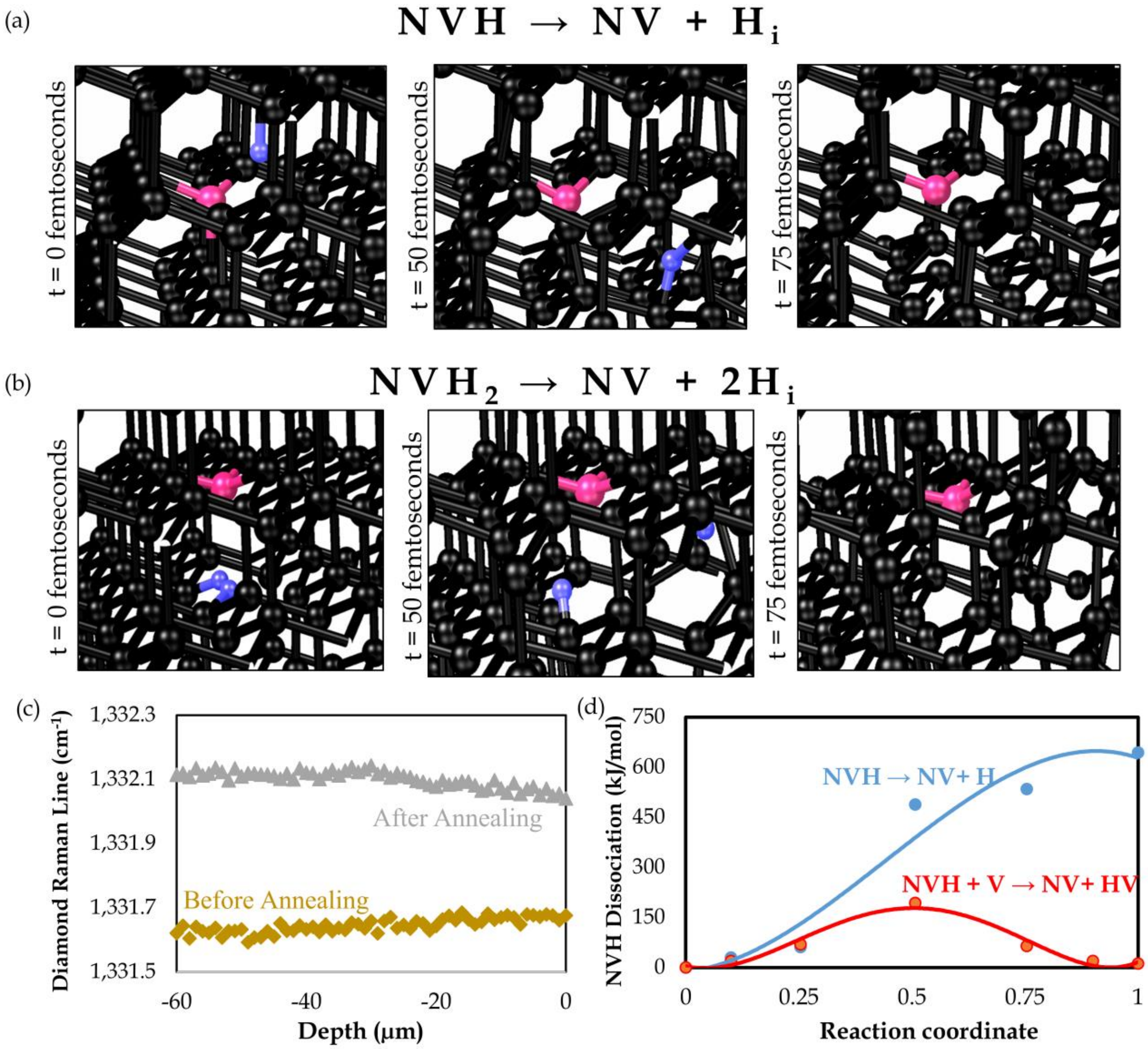

***Fig. 6.*** *The DFT-assisted molecular dynamics results on the dissociation of (a) NVH and (b) $NVH_2$ are shown. In both cases a gradually increasing energy was introduced into the system and upon sufficient energy input, the dissociation of these H-passivated centers into NV was seen in as little as 75 femtoseconds. (c) The diamond Raman line also shows relaxation of the tensile strain towards upon experimental high temperature annealing at 1750 ℃, consistent with our expected lattice relaxation upon hydrogen release. (d) The positive impact of irradiation is also seen through the calculation of energy barriers for the dissociation of NVH through DFT, and we find presence of vacancy at adjacent sites of NVH catalyzes the NV formation drastically by reducing the energy barrier for hydrogen escape from the NV trap.*

The findings explain the promising results on NV formations through femtosecond lasers [61]. Focused femtosecond laser pulses have been used to convert NVH complexes to NV centers in situ, without any furnace anneal. Shimotsuma et al. [62] report that a single 130 fs laser pulse (800 nm, ~nJ energy) can create $NV^-$ centers in diamond without any subsequent thermal treatment, even in bulk samples. The intense, ultrafast laser causes a localized temperature spike and electronic excitation that breaks the C–H bond, thereby freeing an NV center in one step. The breaking of C-H bonds through ultra-fast lasers are shown experimentally in DLC materials recently by Azamoum et al. [63]. Such laser writing results are again consistent with the NVH model: the laser provides the energy to dissociate $NVH_x$ complexes on-demand, as opposed to relying on random vacancy diffusion. More generally, any irradiation (e.g. electron beam, plasma, or laser) that either displaces bonded hydrogen or creates adjacent vacancies can trigger and enhance the conversion efficiency of NVH → NV transformation.

This dissociation mechanism applies broadly to $N_xVH_y$ complexes and provides a unified explanation for multiple experimental observations. In particular, it recontextualizes the known annealing behavior of the 468 nm center, now established as $NVH^-$, which consistently diminishes in intensity during treatments that enhance NV photoluminescence. It also explains the lattice relaxation from tensile strain upon annealing at temperatures between 1750 and 2000 °C, associated with hydrogen release (Fig. 6c). To quantify the transformations and explain the positive impact of electron irradiation on NV formation, we examined the energetics of hydrogen escape from NVH in two scenarios: (i) in a pristine supercell and (ii) in the presence of a nearby irradiation-induced vacancy. In both cases, the energy barrier for the most favorable reaction pathway corresponding to NVH → NV + $H_i$ and NVH → NV + HV were estimated (Fig. 6d). The calculated energy barrier is substantially lowered – by approximately 70% – in the presence of a nearby vacancy, and equaled a hydrogen migration barrier of 1.9 eV, which is lower than the migration energy of vacancies, reported as approximately 2.3 eV [58] highlighting the synergistic effect of irradiation in facilitating hydrogen release and, consequently, NV formation. These findings directly support the observed enhancement of NV conversion efficiency via irradiation–annealing protocols that are prevalent in the literature [14]. Furthermore, this hydrogen-mediated mechanism also explains the irreversible removal of brown coloration upon high-temperature annealing. As previously shown, the brown hue and associated UV–VIS absorption bands at 270, 360, and 520 nm arise from charge transitions of optically active $N_xVH_y$ defects. Heat treatment dissociates these complexes by liberating hydrogen, thereby diminishing their optical activity and reducing tensile strain. The concurrent attenuation of these absorption features, reduction in the 468 nm photoluminescence, and recovery of strain-free Raman signatures confirm this defect-driven model of brown color evolution and NV formation.

These results establish $NVH_x$ dissociation, complementary to vacancy trapping of C-centers, as a dominant pathway for NV center formation in irradiated and annealed CVD diamond. This insight redefines current understanding of NV formation pathway and provides a predictive framework for engineering NV conversion through defect thermodynamics, hydrogen chemistry, and controlled annealing protocols. By linking the evolution of color centers, strain, and NV yield through a single defect transformation mechanism, we provide a comprehensive atomistic model that unifies previously disparate observations in the synthesis of quantum-grade diamond.

**Conclusions**

The controlled formation of nitrogen–vacancy (NV) centers is critical for advancing diamond-based quantum technologies, yet their atomistic formation pathway during irradiation and annealing in CVD grown diamonds remains unresolved. Here, we reveal that hydrogenated NV complexes ($NVH_x$), particularly NVH, $NVH_2$ and $NVH_3$, form readily under growth conditions due to their low formation energies and dominate the defect landscape in as-grown diamond. Using hybrid DFT and spectroscopy, we show that $NVH_x$ complexes are not passive byproducts but thermodynamic endpoints that later dissociate into NV centers during post-growth annealing. This dissociation pathway—driven by thermal activation and electron irradiation—offers a parallel to conventional vacancy-nitrogen pairing. These findings establish $NVH_x$ dissociation as a primary route to NV center formation because of the H migration barrier being lower than the vacancy migration barrier energy and dominance of $NVH_x$ defects during growth, which reframes the role of hydrogen from a complicating impurity to a functional component in quantum defect engineering. The role of $NVH_x$ during growth and their dissociations during post treatment irradiation and annealing unifies many otherwise disparate observations in CVD diamond engineering. It offers a complementary pathway to the conventional vacancy–nitrogen pairing model: instead of creating NV centers from scratch by diffusing vacancies to nitrogen, one can also "unlock" pre-existing NV centers by stripping away hydrogen.

Rather than invoking separate causes for color changes, strain relaxation, and NV yield, the $NVH_x$ dissociation model ties them to a single defect-driven mechanism: the transformation of hydrogenated vacancy defects into un-passivated NV centers. We identify the 468 nm center as a bound exciton transition of $NVH^-$, supported by theoretical lineshapes, FTIR correlations, and Fermi-level-sensitive PL behavior. We additionally demonstrate that the UV–VIS absorption spectrum of brown CVD diamond—including broad bands at 270, 350, and 520 nm—can be accurately reproduced by considering the dielectric response of an ensemble of $N_xVH_y$ defects in thermodynamic equilibrium and the removal of the brown color is explained through dissociations of the hydrogenated nitrogen-vacancy defects. These findings suggest that brown coloration, 468 nm luminescence, and NV formation are not independent phenomena but interconnected stages in the defect transformations of $N_xVH_y$ family.

## Methods

### 3.1 Computational methods

#### 3.1.1 Density functional theory

We have computed the relaxed ground and excited structures, formation energies, Kohn-Sham levels, and potential energy surfaces of N, NV, NVH, $NVH_2$, $NVH_3$, $N_2VH$, and $N_2VH_2$ point defects in diamond, inspired by prior calculations of Feng et al [38]. N represents substitutional nitrogen, V is an adjacent carbon-vacancy, and H are hydrogen atoms bonded to carbon dangling bonds. Periodic boundary conditions are applied with a supercell size of 4×4×4 (512 atoms in the case of a pristine cell), which is large enough to avoid any spurious defect-defect interactions [64, 65]. Hybrid spin-polarized density functional theory is applied by using the projector augmented wave method (PAW) [66, 67]. The Heyd-Scuseria-Ernzerhof (HSE06) [68, 69] exchange functional is used, with a k-point spacing of 0.44 $Å^{-1}$, which translates to a Γ-sampling. Calculations are carried out using reciprocal space with a plane-wave cutoff-energy of 520 eV. The Gaussian integration scheme is applied, with a smearing width of 0.05 eV. During atomic relaxations, the calculations (under constant volume) are iterated until atomic positions are such that the Hellmann-Feynman forces are smaller than 10 meV/Å, with a self-consistency convergence threshold of $10^{-6}$ eV. The calculations are conducted through the VASP code and MedeA environment.

#### 3.1.2 Defect formation energies and concentrations

The defect formation energies ($\Delta H_f^q$) are calculated using Eq.1 [70].

$$\Delta H_f^q = E_{tot}^q - \sum n_i \mu_i + q(E_f + E_{VBM}) + E_{corr}^{FNV} \quad \text{(Eq.1)}$$

$E_{tot}^q$ is the difference between the calculated supercell energies of a defective supercell of charge $q$, and the pristine supercell ($E^{Defect}$ - $E^{Pure}$). $\mu$ is the chemical potential of an added or removed element for producing the defective supercell, and $n$ is the stoichiometric coefficient. In other words, $n$ is +1 (-1) if an atom of $\mu_x$ chemical potential is added (removed). The hydrogen and nitrogen chemical potentials are referenced to the $H_2$ and $N_2$ molecules, respectively. Carbon chemical potential is retrieved from the bulk diamond. $E_f$ is the Fermi energy of the diamond crystal referenced to valence band maximum ($E_{VBM}$). $E_{corr}$ is the correction for finite supercell size and spurious electrostatic interactions, which is inevitable in the calculation of charged defects [70]. We have applied the fully *ab initio* correction scheme proposed by Freysoldt, Neugebauer, and Van de Walle (FNV) [71], through the SXDEFECTALIGN code. Formation energies are plotted as a function of the Fermi energy.

The relative concentrations of defects in a crystal depend on the equilibrium Fermi energy, due to the Fermi energy dependence of formation energies (Eq. 1). Numerous experimental works on as-grown CVD diamond have found that the concentrations of $N^{1+}$ and $N^0$ are within the same order of magnitude, with various reports regarding the slight dominance of one over the other. This means that the equilibrium Fermi energy must be pinned at or very close to the N (+/0) thermodynamic charge transition, which is at $E_{VBM}$+3.45 eV (Section 3.1). Therefore, the concentration of each defect ([$d$]) has been computed using Eq. 2 [70], by considering $E_f$ values of 3.3, 3.4, and 3.5 eV in Eq. 1. Temperature effects are applied based on changes in chemical potentials retrieved from the JANAF thermochemical database, as explained in detail elsewhere [24, 72].

$$[d]_i = N_{d_i}\, exp\left(\frac{-\Delta H_{i_f}^{q}}{kT}\right) \qquad \text{(Eq. 2)}$$

Here, $N_{di}$ is the number of sites the defect can occupy within the pristine diamond cell per unit volume. $\Delta H_{i_f}^{q}$ is the formation energy of defect for a given Fermi energy (Eq.1), $K$ is the Boltzmann constant, and $T$ is the growth temperature, which is considered as 950 °C. The calculation of the equilibrium Fermi energy and temperature dependent changes in the defect formation energies as well as the charge neutrality were carried out similar to the multiscale methodology published earlier and applied on GaN nd AlN [73, 74].

### 3.1.3 Charge transition levels and configuration coordinate diagrams

The thermodynamic charge transition levels correspond to a Fermi energy where two different charged species of a defect have equivalent formation energy. Therefore, the adiabatic transition levels for charges $q_1/q_2$ of a defect are expressed as follows (Eq. 3).

$$\varepsilon^{q_1/q_2} = \frac{\Delta H_f^{q_1}(E_f = 0) - \Delta H_f^{q_2}(E_f = 0)}{q_2 - q_1} \qquad \text{(Eq. 3)}$$

The transition levels $\varepsilon^{q_1/q_2}$ and $\varepsilon^{q_1/q_2}$ correspond to zero phonon lines (ZPL) for electron and hole captures respectively, where $q_1 < q_2$. The peak absorption energy can be estimated using the Franck Condon principle. Therefore, we have plotted configuration coordinate diagrams (CCD) for the expected NVH transitions. The CCD are calculated through estimation of the one dimensional potential energy surface for the ground and excited charged states, as proposed by Alkauskas et al. [75, 76]. We have calculated the formation energies for 9 geometries using hybrid DFT (HSE06) parameters mentioned earlier. The change in configuration coordinates ($\Delta Q$) for the excited state is calculated using Eq. 4, where $\Delta R$ is defined by Eq.5. The ground state configuration coordinate is taken as Q = 0 amu$^{1/2}$Å. The success of this methodology has been demonstrated in a wide range of semiconducting and insulator materials.

$$(\Delta Q)^2 = \sum_{i,j} m_i \Delta R_{i,j}^2 \qquad \text{(Eq. 4)}$$

$$\Delta R = R_{ex} - R_{gr} \quad \text{(Eq. 5)}$$

$R_{ex}$ and $R_{gr}$ refer to the excited and ground state atomic coordinates, respectively.

### 3.1.4 Theoretical absorption spectra and color

Due to the wide band gap of diamond, a pristine supercell that is free of any defects must be inherently transparent in the visible spectrum. However, defects can cause a range of colors based on their internal and charge transition levels, concentrations, respective extinction coefficients, material thickness, and illumination light source. To compute the absorption spectra, and consequently the color caused by the nitrogen related defects considered in this study, we have calculated the real and complex dielectric functions for each defective supercell through the HSE06 level of theory as mentioned in section 2.1.1. We have applied the methodology of Gajdoš et al. [77] for the calculation of the imaginary ($\epsilon^i$) and real ($\epsilon^r$) parts of the dielectric function as a function of photon frequency $\omega$ (Eq.6, 7).

$$\epsilon^i(\omega) = \left(\frac{4\pi^2 e^2}{\Omega}\right) \lim_{\mathcal{L}\to 0} \frac{1}{\mathcal{L}^2} \sum_{c,v} 2\delta(\epsilon_c - \epsilon_v - \omega) \times \langle u_{c+e_\alpha \mathcal{L}} \mid u_v \rangle \langle u_{c+e_\beta \mathcal{L}} \mid u_v \rangle^* \quad \text{(Eq.6)}$$

$\Omega$ is the supercell volume. The conduction and valence band states are represented by $c$ and $v$ indices, respectively. $\mathcal{L}$ is the Bloch vector of the incident wave, and $u_c$ is the cell-periodic part of the orbitals at the Γ-point. The real component of the dielectric tensor is then obtained through Kramers-Kroning transformation as follows, where P is the principal value, and $\eta$ is the complex shift taken as 0.1 which serves the purpose of transformation smoothening (Eq. 7). For a detailed treatment of this approach, see the prior work by Gajdoš et al. [77].

$$\epsilon^r(\omega) = 1 + \left(\frac{2P}{\pi}\right) \int_0^\infty \frac{\mathcal{E}^i(\omega')\omega'}{\omega'^2 - \omega^2 + i\eta} d\omega' \quad \text{(Eq. 7)}$$

The optical frequency dependent absorption coefficient ($\Lambda$) induced by defect $i$ can be finally calculated using Eq. 8, where c is the speed of light in vacuum [78].

$$\Lambda_i(\omega) = \frac{\sqrt{2}\omega}{c}\sqrt{\sqrt{\epsilon^{r2} + \epsilon^{i2}} - \epsilon^r} \qquad \text{(Eq. 8)}$$

Each defect will have a distinct $\Lambda_i(\omega)$ with respect to a given defect concentration and will therefore have a partial contribution towards the total absorption coefficient, $\Lambda_T(\omega)$. We have formulated a partition function (Eq. 9) based on the concentration of each defect to achieve the absorption coefficient for the entire system of nitrogen related defects studied in this work.

$$\Lambda_T(\omega) = [C_N]\sum_{i=0}^{n} \Lambda_i(\omega)\left(\frac{[d]_i}{\sum_{j=0}^{n}[d]_j}\right) \qquad \text{(Eq. 9)}$$

$[C_N]$ represents the total nitrogen concentration in a diamond crystal, and *n* is the total number of considered defects (n=19 defects, as each charged defect has a distinct contribution to the partition function).

To evaluate the color that can arise from $\Lambda_T(\omega)$, the spectrum is then converted into transmission spectrum $T(\omega)$. This is possible through Bouguer-Lambert's law (Eq. 10), where $L$ is the path-length of light in the crystal (effective thickness).

$$T(\omega) = e^{-L\,\Lambda_T(\omega)} \qquad \text{(Eq. 10)}$$

Finally, the respective standard red green blue (sRGB) color coordinates are evaluated based on the $D_{65}$ illumination reference, through conversion of the transmission spectra into XYZ tristimulus values, which range from 0 to 1 (Eq. 11-13) [79, 80].

$$X = \frac{1}{\xi}\int \bar{x}(\omega)\, T(\omega)\, I(\omega)\, d\omega \qquad \text{(Eq. 11)}$$

$$Y = \frac{1}{\xi} \int \bar{y}(\omega)\, T(\omega)\, I(\omega)\, d\omega \qquad \text{(Eq. 12)}$$

$$Z = \frac{1}{\xi} \int \bar{z}(\omega)\, T(\omega)\, I(\omega)\, d\omega \qquad \text{(Eq. 13)}$$

$I(\omega)$ is the spectral power distribution of the illuminant and thus depends on the light source. We have used the CIE standard $D_{65}$ illuminant, which corresponds to the average midday sunlight in Western Europe. $\bar{x}(\omega)$, $\bar{y}(\omega)$, and $\bar{z}(\omega)$ are the CIE standard observer functions. The integration for X, Y, and Z are carried out for the visible spectrum range. The value of $\xi$ is obtained through Eq. 14.

$$\xi = \int \bar{y}(\omega)\, I(\omega)\, d\omega \qquad \text{(Eq. 14)}$$

The XYZ tristimulus is then converted to sRGB through the following transformation matrix (Eq. 15).

$$\begin{bmatrix} R \\ G \\ B \end{bmatrix} = 255 \cdot \begin{bmatrix} 3.240 & -1.537 & -0.499 \\ -0.969 & 1.876 & 0.042 \\ 0.056 & -0.204 & 1.057 \end{bmatrix} \begin{bmatrix} X \\ Y \\ Z \end{bmatrix} \qquad \text{(Eq. 15)}$$

In other words, the calculated RGB values depend on the total nitrogen content, defect concentrations (which depend on the Fermi energy and growth conditions), illumination source, and effective path length of light in a diamond. We have used the GemRay software for illustration of the calculated colors through rendering a diamond with the facet diagram shown in Fig. 1 that has an effective light path length of 4mm.

### 3.1.5 Vibrational frequencies and photoluminescence spectra

The phonon spectra are calculated using density functional perturbation theory (DFPT) [81] by applying the generalized gradient approximation (GGA) [82], as parametrized by Perdew-Burke-Ernzerhof (PBE). The 4×4×4 supercell is used for all calculations, with a planewave cut-off energy of 620 eV. A strict force convergence of $10^{-4}$ eV/Å and self-consistency criteria of $10^{-6}$ eV are applied. These parameters reproducer experimental results in diamond with a remarkable level of accuracy as shown by various prior studies in the literature [64, 81].

Simulating the photoluminescence (PL) line-shape induced by a defect requires knowledge of the vibronic coupling. This is the coupling between vibrational and electronic states of the excited and ground state structures. Therefore, the PL line-shape $I\ (\hbar\omega)$ at zero Kelvin is described as follows (Eq. 16).

$$I\ (\hbar\omega) = I_0 \sum_{m}^{M} \left|\langle\psi_g \mid \psi_e\rangle\right|^2 \delta\ (E_{ZPL} - E_{gm} - \hbar\omega) \qquad \text{(Eq. 16)}$$

where M is the total number of levels, $E_{ZPL}$ is the energy of the zero-phonon line, and $E_m$ is the energy of the vibrational band $m$. $\psi_g$ and $\psi_e$ are the ground and excited state wavefunctions. Computing PL line-shape through Eq. 16 requires the solution of many-body Schrödinger's equation, which is non-trivial. Therefore, we have applied a methodology developed by Alkauskas et al. [81], and its subsequent implementation in the PyPhotonics code [83]. In this approach, the normalized luminescence intensity $L\ (\hbar\omega)$ is computed, rather than the absolute intensity $I\ (\hbar\omega)$, as demonstrated in Eq. 17.

$$L\ (\hbar\omega) = C\omega^3 \sum_{m}^{M} \left|\langle\chi_g \mid \chi_e\rangle\right|^2 \delta\ (E_{ZPL} - E_{gm} - \hbar\omega) \qquad \text{(Eq. 17)}$$

Here, $\chi_g$ and $\chi_e$ are the ground and excited state vibrational levels. $E_{gm}$ and $C$ are defined as follows (Eq. 18 and 19).

$$E_{gm} = \sum_{m}^{M} n_m \hbar\omega_m \qquad \text{(Eq. 18)}$$

$$C^{-1} = \int \omega^3 \sum_{m}^{M} \left| \langle \chi_g \mid \chi_e \rangle \right|^2 \delta \left( E_{ZPL} - E_{gm} - \hbar\omega \right) d(\hbar\omega) \qquad \text{Eq. 19}$$

For an in-depth treatment of this methodology refer to the prior work by Alkauskas et al. [81]. We have performed a benchmarking on the $NV^-$ center using the parameters applied in this study, and the results are in excellent agreement with the prior theoretical calculations, and experimental data as shown in Appendix A.

### 3.1.6 Calculation of the migration barrier and molecular dynamics simulations

The migration barriers and nudge elastic calculations [84] were carried out through VASP 6.1 and the Transient State Search module of MedeA. The generalized gradient approximations were used as the exchange functional. A 256 atom supercell was used, with the two cases of hydrogen migration: H-migration to an interstitial site from NVH, leaving behind an NV center, and H-migration from NVH to an adjacent vacancy leaving behind an NV center.

To simulate the behavior of $NVH_x$ centers upon energy input (radiation, heat etc.), we used spin polarized DFT for energy and force calculations for each frame of the dynamical calculations, and applied temperature scaling (nVE) ensemble, as explained in detail by Zhou et al. [85]. A simulation time of 100 fs and a time step of 1 fs was used with the initial temperature of 0 K to a final temperature of 2500 K.

## 3.2 Experimental Methods

Two sets of diamond samples were used, grown the GIA and Appsilon BV. The growth conditions under microwave plasma varied, but principally were 200 sccm $H_2$, 5-10 sccm $CH_4$, 0-10 sccm $O_2$, none to minimal $N_2$ flow. The total pressure of 300 mbar was used and a temperature of approximately 920 - 1000 °C, MW power of ~4000 W; and a growth rate of ~25 micron/hour.

Fluorescence and phosphorescence imaging (figure 2m) was collected with a DiamondView microscope which uses a pulsed xenon flash lamp providing near and above bandgap excitation ( < 225 nm), as described earlier [86]. The IR mapping (figure 4a) was conducted using a Nicolet iN10 IR mapping instrument. The IR mapping technique is best performed with flat plates and collecting the transmission absorption spectra. The detector was cooled by liquid nitrogen. Each spectrum in the map includes 64 scans over the mid-infrared range (6000 to 675 $cm^{-1}$) with a resolution of 1 $cm^{-1}$ and an aperture size of 300 µm x 50 µm. The IR spectra were extracted from the collected map and each was background subtracted and normalized using the two-phonon absorption of diamond [87]. Finally, the integrated area above baseline was determined for the 3123 $cm^{-1}$ $NVH^0$ peak.

PL mapping (figures 2n, 2o, 3b, 4a) was obtained using a Thermo Fisher Scientific DXR2xi. The PL spectral maps were collected at liquid nitrogen temperature with either 455, 532, or 785 nm excitation as appropriate, 10× magnification, and shown in units of wavelength (nm). As PL spectra are semi-quantitative, the PL peak areas determined from the spectra were normalized by using the calculated integrated area of the unsaturated diamond Raman peak. Cross-sectional TEM lamellas (with thickness ~ 100 nm) were prepared from diamond samples with Aquilos2 focused ion beam (FIB) system using 30 keV Ga ions and cleaned using 2 keV FIB. The TEM transparent areas on each prepared lamella was around 40-50 nm. Bright field TEM, HAADF STEM imaging and EELS were conducted using Tecnai F30 microscope operated at 200 keV. EELS measurements were conducted in STEM mode, with 1 nm probe beam diameter, convergence semi-angle of 2.2 mrad and collection semi-angle of 16 mrad using Gatan GIF QuantumTM 965 energy filter with dual EELS capability.

Secondary ion mass spectrometry (SIMS) analyses were conducted using a CAMECA IMS 1280 instrument housed at the WiscSIMS Laboratory, University of Wisconsin. A primary $Cs^+$ ion beam with a current of 2.5 nA was focused onto the sample surface with an oval-shaped spot size of approximately 10 × 12 $\mu m^2$. Simultaneous detection of secondary ions corresponding to $^{24}[^{12}C^{12}C]$ and $^{26}[^{12}C^{14}N]$ was achieved using two Faraday cups. Further details regarding the analytical setup are provided in reference [88]. Quantification of nitrogen concentrations was calibrated using diamond standards with previously characterized nitrogen content, determined by SIMS at the Canadian Center for Isotopic Microanalysis.

## CRediT Statement

**Mubashir Mansoor:** Conceptualization, Methodology, Software, Data Curation, Investigation, Visualization, Formal Analysis, Resources, Writing-original draft, Funding Acquisition, Project Management. **Kamil Czelej:** Supervision, Resources, Conceptualization, Methodology, Software, Investigation, Data Curation, Investigation, Formal Analysis and Writing-original draft. **Sally Eaton-Magaña:** Spectroscopic Methodology, Validation, Investigation, Data Curation, Investigation, formal analysis Writing-review & editing. **Mehya Mansoor:** Conceptualization, Methodology, Software, Data Curation, Investigation, Formal Analysis and Writing-original draft, Visualization. **Rümeysa Salci**: Validation, Investigation, Data Curation, Formal Analysis, Visualization, Writing-review & editing. **Maryam Mansoor:** Conceptualization, Methodology, Software, Data Curation, Investigation, Formal Analysis and Writing-original draft, Visualization. **Taryn Linzmeyer:** Investigation, Data Curation, formal analysis, Writing-review & editing, Visualization. **Yahya Sorkhe:** Investigation, Data Curation, Formal Analysis, Visualization, Writing-review & editing. **Kyaw S. Moe:** Data Curation, Investigation. **Ömer Özyildirim:** Software, Investigation, Data Curation, Methodology. **Kouki Kitajima:** Investigation, Data Curation, Formal Analysis. **Mehmet Ali Sarsil:** Investigation, Data Curation, Visualization, Formal Analysis, Software. **Taylan Erol:** Resources, Investigation, Funding Acquisition, Formal Analysis. **Gökay Hamamci**: Funding Acquisition, Resources. **Onur Ergen**: Funding Acquisition, Formal Analysis, Resources, Supervision. **Adnan Kurt:** Formal Analysis, Supervision. **Arya Andre Akhavan:**,Formal Analysis, Visualization, Software, Investigation, Data Curation. **Zuhal Er**: Supervision, writing-original draft, Methodology, Software. **Sergei Rubanov:** Methodology, Investigation, Data Curation, Formal Analysis, Visualization. **Nikolai M. Kazuchits:** Data Curation, Formal Analysis, Investigation. **Aisha Gokce:** Resources, Supervision. **Nick Davies:** Formal Analysis, Investigation, Data Curation, Visualization. **Servet Timur:** Supervision, Writing-review & editing, Formal Analysis, Resources, Methodology. **Steven Prawer:** Supervision, Conceptualization, Methodology, Formal Analysis, Data Curation, Resources, Investigation, Writing-review & editing. **Alexander Zaitsev:** Supervision, Conceptualization, Methodology, Formal Analysis, Data Curation, Resources, Investigation, Writing-original draft, Visualization. **Mustafa Ürgen:** Supervision, Funding Acquisition, Conceptualization, Methodology, Formal Analysis, Data Curation, Resources, Investigation, Writing-original draft, Project Management.

**Acknowledgements**

Part of this research was funded by the European Research Council (QUEEN project: 01043219) and Quantum Delta Nederland (SME_R8_QuMat), which are acknowledged. WiscSIMS is supported by NSF (EAR-2320078) and the University of Wisconsin-Madison. Computing resources were provided by High Performance Computing facilities of the Interdisciplinary Centre for Mathematical and Computational Modeling (ICM) of the University of Warsaw under Grant No. GB79–16 and Poznan Supercomputing and Networking Center (PSNC) as well as the National Center for High Performance Computing of the republic of Türkiye (UHeM) under grant number 1008852020.

---

# Supplementary information

## 1. Defect equilibria and growth conditions

The $N_xVH_y$ defect formation energies depend on the chemical potentials of nitrogen ($\mu_N$) and hydrogen ($\mu_H$) during growth, as presented in Eq. S1. [A1, A2]

$$\Delta H_f^q = E_{tot}^q - \sum n_i \mu_i + q(E_f + E_{VBM}) + E_{corr}^{FNV} \quad \text{(Eq. S1)}$$

In other words, the total concentration in any given point defect ensemble depends on the formation energies of the individual defects, which are dictated by the chemical potential of a defect's constituent elements. Therefore, the growth conditions gain significant importance as the $\mu$ values directly depend on the growth conditions, such as temperature, partial pressures of gases, absolute pressure, use of plasma or lack thereof, temperature of the plasma, distance from the plasma core to substrates and so on. Each of these parameters directly influence the defects that form in diamond and their relative concentrations are inevitably affected. To further explore the changes in the chemical potential equilibria for the case of hydrogen and nitrogen under a plasma environment, a CALPHAD calculation [A3] was done on the thermodynamic activities of the H-C-N databases for finding the various dominant nitrogen and hydrogen related species, as a function of pressure and plasma temperature. The rising thermodynamic activities ($a_i$) shown in Fig. S1 demonstrate an increasing chemical potential ($\mu_i$), based on the relation shown in Eq. S2, where $k$ is the Boltzmann constant, $T$ is temperature and ${\mu_i}^0$ is the standard reference chemical potential of specie $i$

$$\mu_i(T,P) = E_{ref} + \mu_i{}^0(T) + kTln(a_i) \qquad \text{(Eq. S2)}$$

The CALPHAD calculations reveal an increasing H activity under reduced total pressures and increasing temperatures, primarily due to a more effective dissociation of $H_2$ into atomic H (Fig. S1). Moreover, in the case of nitrogen, the most dominant species are found to be $N_2$ and HCN, based on the pressure and plasma temperature. Therefore, considering a plasma ball within a MWCVD reactor, the nonhomogeneous plasma on the diamond seeds, will expose the various diamond seeds to different localized chemical potential equilibria, which in turn result in different colorations, various shades of brown (upon N-doping), and from an atomistic point of view; different point defect ensembles.

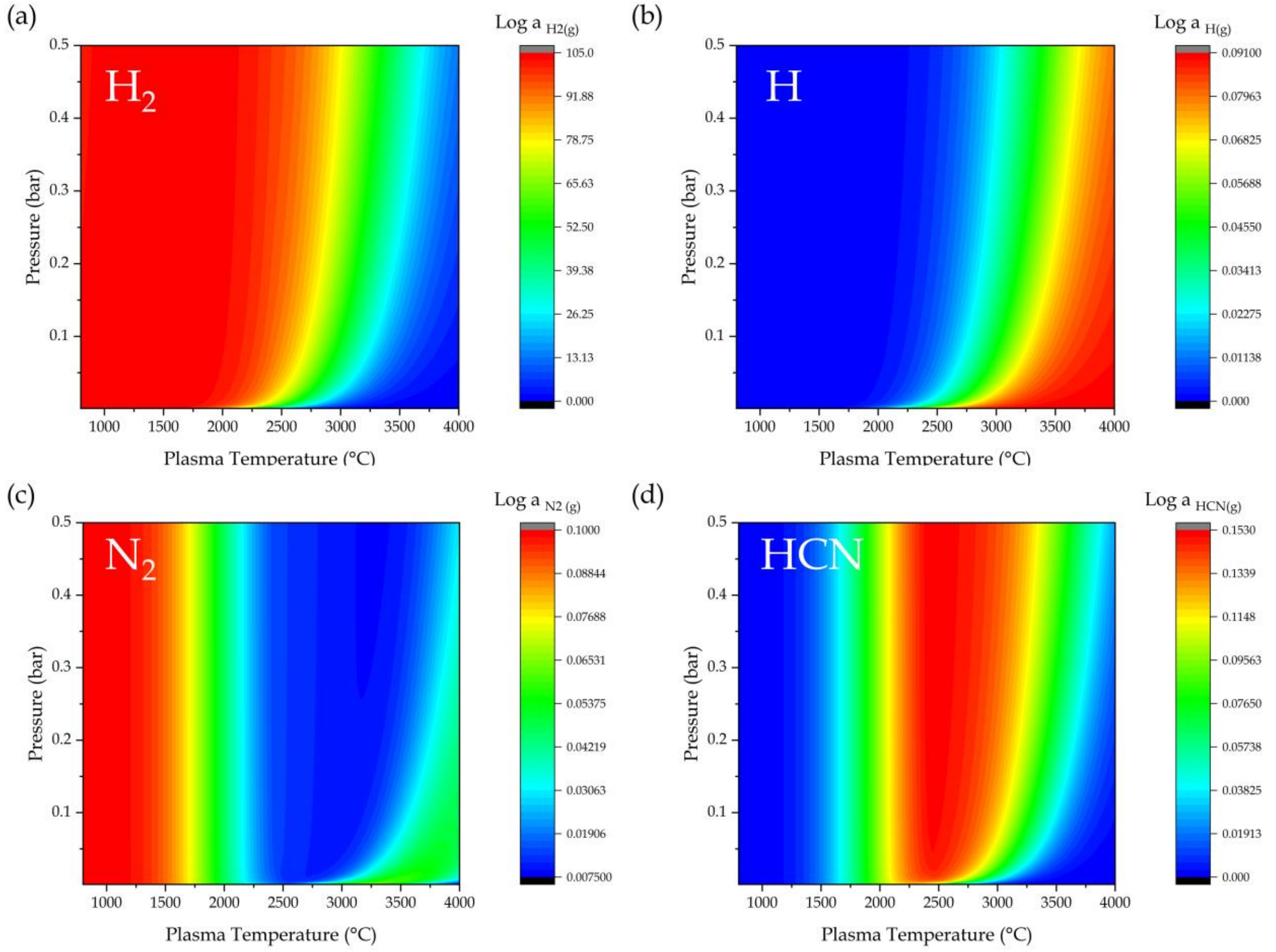

***Fig S1.*** *The activity of dominant hydrogen and nitrogen related species are demonstrated as a function of plasma temperature and pressure for (a)* $H_2$*, (b) H, (c)* $N_2$*, and (d) HCN*

The formation energies of defects are directly influenced by the parameters discussed earlier. To better illustrate these influences, we focus here on the effect of temperature, highlighting the magnitude and significance of changes in defect formation energies—changes that render assumptions based solely on zero Kelvin energies insufficient for capturing experimental observations (Fig. S2). The temperature effect is introduced via Eq. S2, which reflects a shift in the chemical potential of species involved in defect formation. A detailed description of the methodology is available elsewhere [A4, A5].

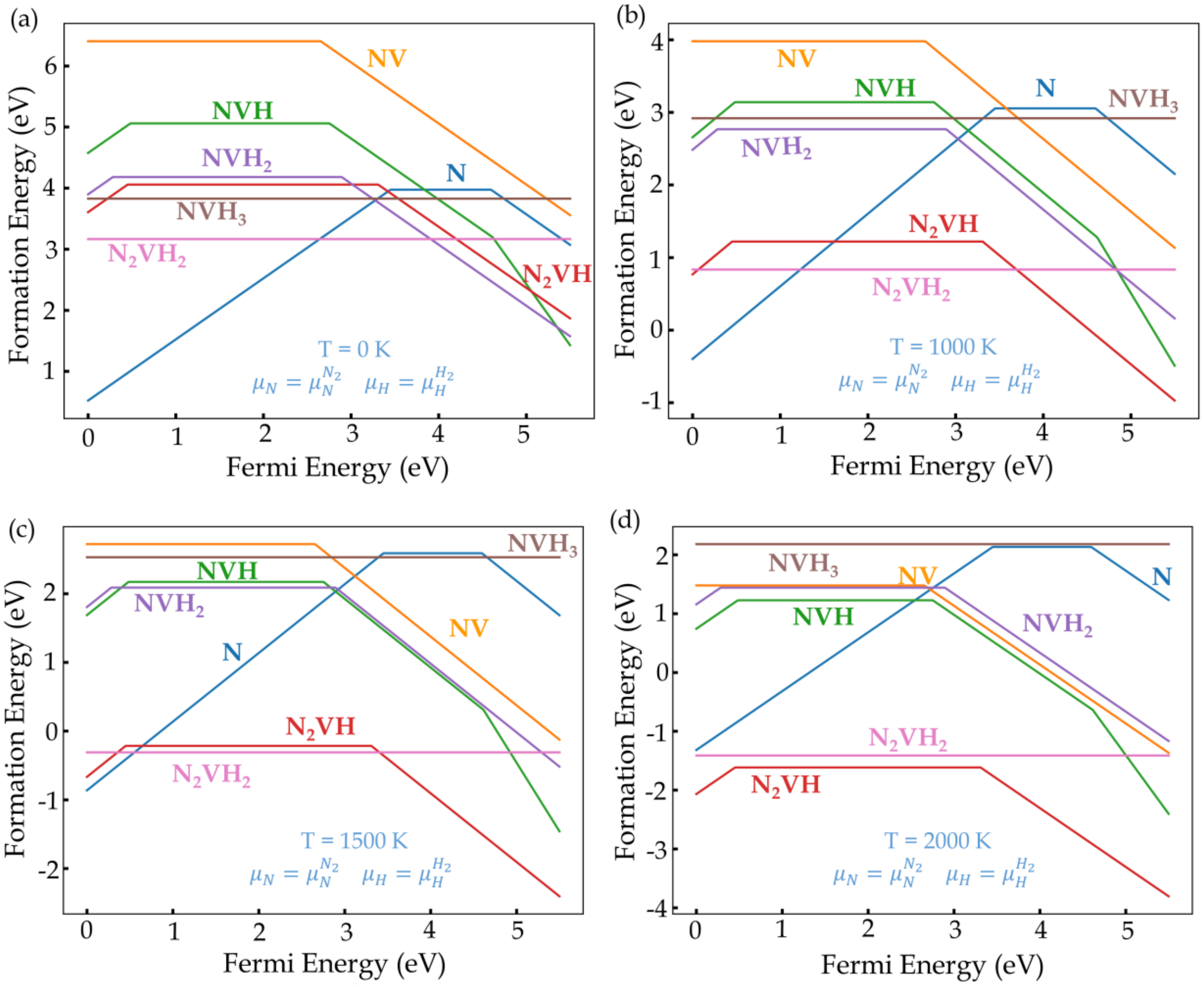

***Fig S2.*** *The activity of dominant hydrogen and nitrogen related species are demonstrated as a function of plasma temperature and pressure for (a) $H_2$, (b) H, (c) $N_2$, and (d) HCN*

As the substrate temperature increases (with all other conditions constant), the defect $N_2VH$ becomes increasingly dominant. However, near 1500 K, the formation energies of both $N_2VH$ and $N_2VH_2$ approach negative values. Practically, a negative formation energy indicates the onset of phase dissociation, setting an upper temperature limit for diamond growth under the presented chemical potential equilibria. This relationship is governed by Eq. S3, which dictates defect concentrations:

$$\frac{[d]_i}{N_d} = exp\left(\frac{-\Delta H_{i_f}^{q}}{kT}\right) \qquad \text{(Eq. S3)}$$

If $\frac{[d]_i}{N_d} > 1$, it follows that defect concentration exceeds available diamond lattice sites, which from a thermodynamic stand point leads to instability and dissociation of the diamond phase. This is inevitable once $\Delta H_{i\,f}^{\,q} < 0$.

To calculate defect concentrations comprehensively across a wide range of growth conditions, we treat the chemical potentials of nitrogen and hydrogen as variables and solve for the emergent defect equilibria in each case (Fig. S3). This requires the use of charge neutrality within a grand canonical approach—now a standard methodology in defect physics. Compared to the earlier Kroger-Vink method, which demanded complex bookkeeping of each possible defect reaction, this modern approach simplifies the analysis significantly. For further details, see Freysoldt et al. [A6].

Charge neutrality is a foundational principle stating that the total positive and negative charges in a material must sum to zero; any net charge would be non-physical for a bulk solid. In practice, this means that the concentrations of electrons, holes, and all charged defects must balance exactly. Defect concentrations, therefore, cannot be determined independently—they are tightly coupled with the equilibrium Fermi energy, which itself depends on the defect landscape and carrier populations. Solving for charge neutrality ensures a self-consistent determination of both Fermi energy and defect concentrations.

Defect concentrations as functions of process conditions can be visualized through Kröger-Vink diagrams, constructed based on the equilibrium formation energies (Eq. S1) determined from the ensemble's equilibrium Fermi energy. Charge neutrality is the only viable path to obtaining this equilibrium state. Specifically, the system must satisfy the following condition:.

$$\sum_{i=1}^{M} q_i[d]_i + [p] - [n] = 0 \qquad \text{(Eq. S4)}$$

Here $[d]$ is the concentration of a defect $i$ with charge $q_i$. $M$ is the total quantity of defects taken into account for a particular defect ensemble. The concentrations of holes and electrons are denoted by $[p]$ and $[n]$, respectively, which are calculated as follows.

$$[p] = \int_{E_{VBmin}}^{E_{VBmax}} \frac{g_v(E)dE}{1 + \exp\left(\frac{E_f - E}{kT}\right)} \qquad \text{(Eq. S5)}$$

$$[n] = \int_{E_{CBmin}}^{E_{CBmax}} \frac{g_c(E)dE}{1 + \exp\left(\frac{E - E_f}{kT}\right)} \qquad \text{(Eq. S6)}$$

The density of states for the valence and conduction bands, as computed by DFT for the pristine diamond supercell, are denoted by $g_v(E)$ and $g_c(E)$, respectively. Somjit & Yildiz [A5] present a detailed description of the methodology and their work is highly recommended for more information in this regard. Through the implementation of this methodology, the defect equilibria as function of variable growth conditions is achieved, which are presented in Fig. S4.

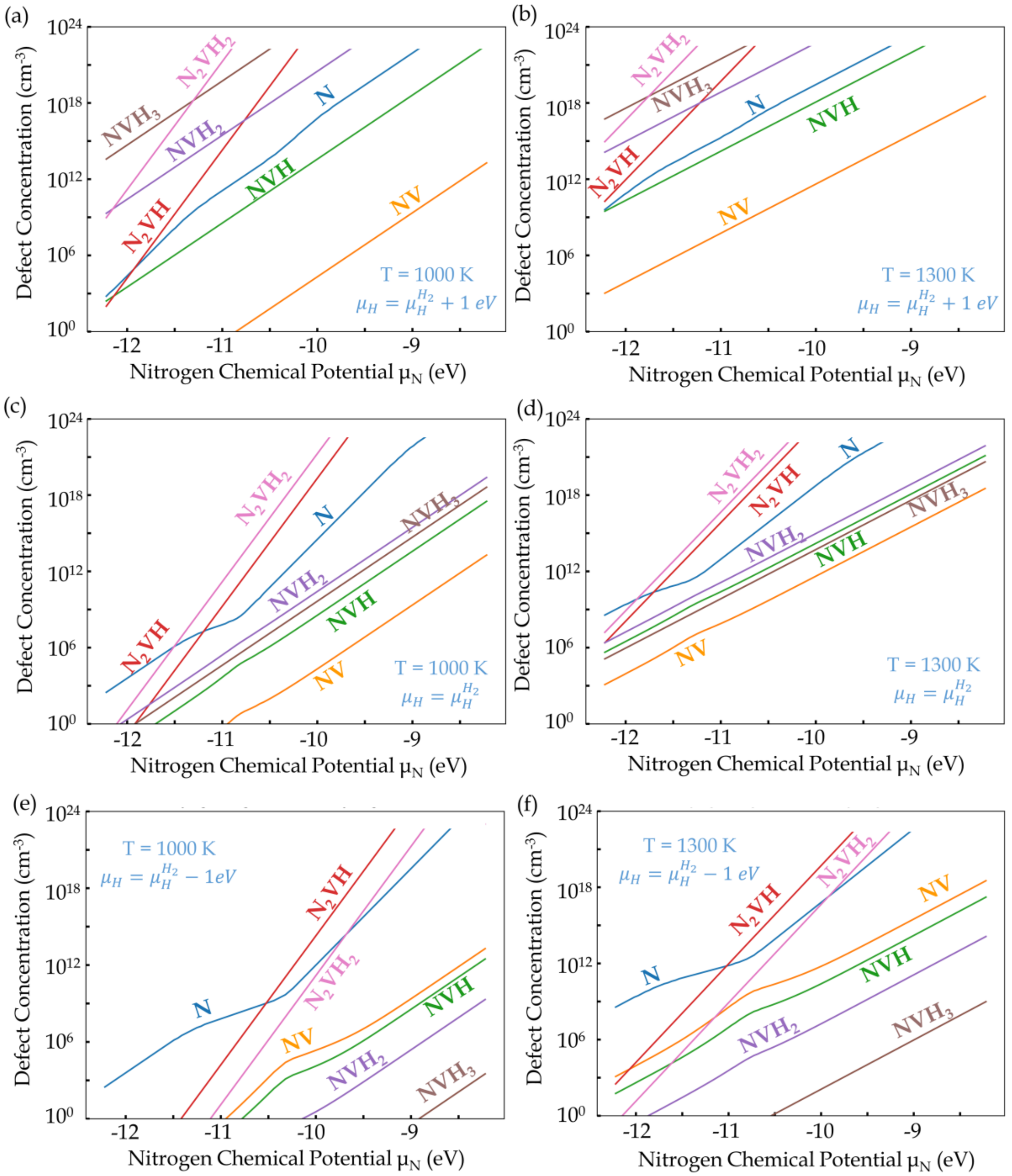

***Fig. S3.*** *The N-V-H defect equilibria as a function of variable hydrogen and nitrogen chemical potential are shown for two different temperatures of 1000 and 1300 K.*

Increasing hydrogen chemical potential, as in microwave plasma, triggers formation of enhanced concentrations of $N_xVH_y$ defects, whereas single substitutional N becomes dominant if the $\mu_H$ is less than that of the hydrogen chemical potential in $H_2$ molecule. Therefore, it is fair to expect significantly less formation of hydrogen passivated centers in hot-filament CVD systems in comparison to MWCVD. The formation of $N_2VH$ and $N_2VH_2$ are highly likely upon most growth conditions. Similarly, upon increase in $\mu_H$ the rise in the concentration of hydrogen passivated NV centers should be expected, and thus there can exist localized regions where NVH:NV ranges from 0.01 to 100, depending on the hydrogen and nitrogen activities at that particular localized growth environment.

## 2. Strain induced by $N_xVH_y$ versus vacancy clusters, TEM observation and electron density

The strain induced by hydrogen and nitrogen related defects has been discussed in the main text, which is inherently tensile and can reach strains of even 1% dilation of the unit cell. The DFT calculations on the impact of strain, irrespective of the straining source, is illustrated in Fig. S4. The Raman 1332 $cm^{-1}$ diamond line deviates to lower $cm^{-1}$ values upon lattice expansion, and the effect of hydrostatic stress on the deviation of the 1332 $cm^{-1}$ is found to be 2.93 GPa per every $cm^{-1}$ change in the diamond Raman line. The tensile strain also causes a reduction in the bandgap of diamond, however, the magnitude is not significant enough to induce any changes in the coloration. The methodology of Parlinsky et al. [A7] is used for Raman calculations.

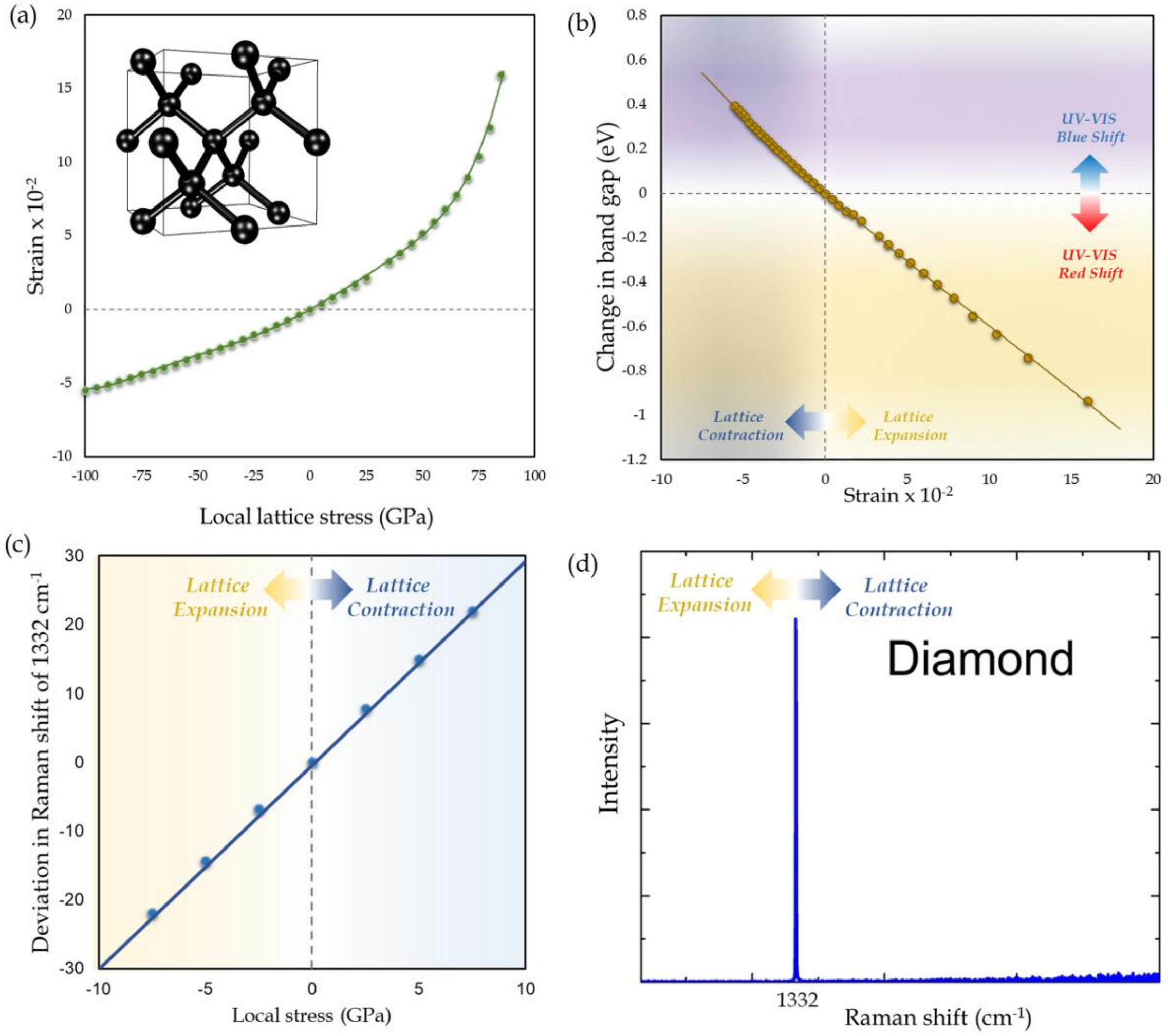

***Fig. S4***. *(a) Strain and the equivalent hydrostatic stress relationship for diamond is shown. (b) Tensile strains cause a reduction of the electronic band gap, and (c,d) deviate the diamond Raman line to lower* $cm^{-1}$ *values.*

The type of strain induced by vacancy clusters has also been investigated, and we found although single carbon vacancies have negligible impact on lattice strain, the impact of vacancy clusters can be quite significant, with the nature of such strain being compressive. A 512 atom diamond cell was used for modelling of the vacancy clusters with sizes of 2, 5, 8, and 16 carbon vacancies. We consistently found that increasing size of a vacancy cluster results in a greater compression of the diamond unit cell, primarily due to relaxation of the carbon atoms at the peripheries of the cluster which shift their Wyckoff coordinates to new equilibrium positions relatively towards the center of vacancy cluster (white atoms in Fig. S5b), thus causing a shrinkage in the lattice parameters. By extrapolating from the DFT calculated data given here (Fig. S5a), a 60-atom vacancy cluster results in a localized compressive stress of 7 GPa, which translates to the diamond Raman line shifting to values between 1333 to 1334 $cm^{-1}$.

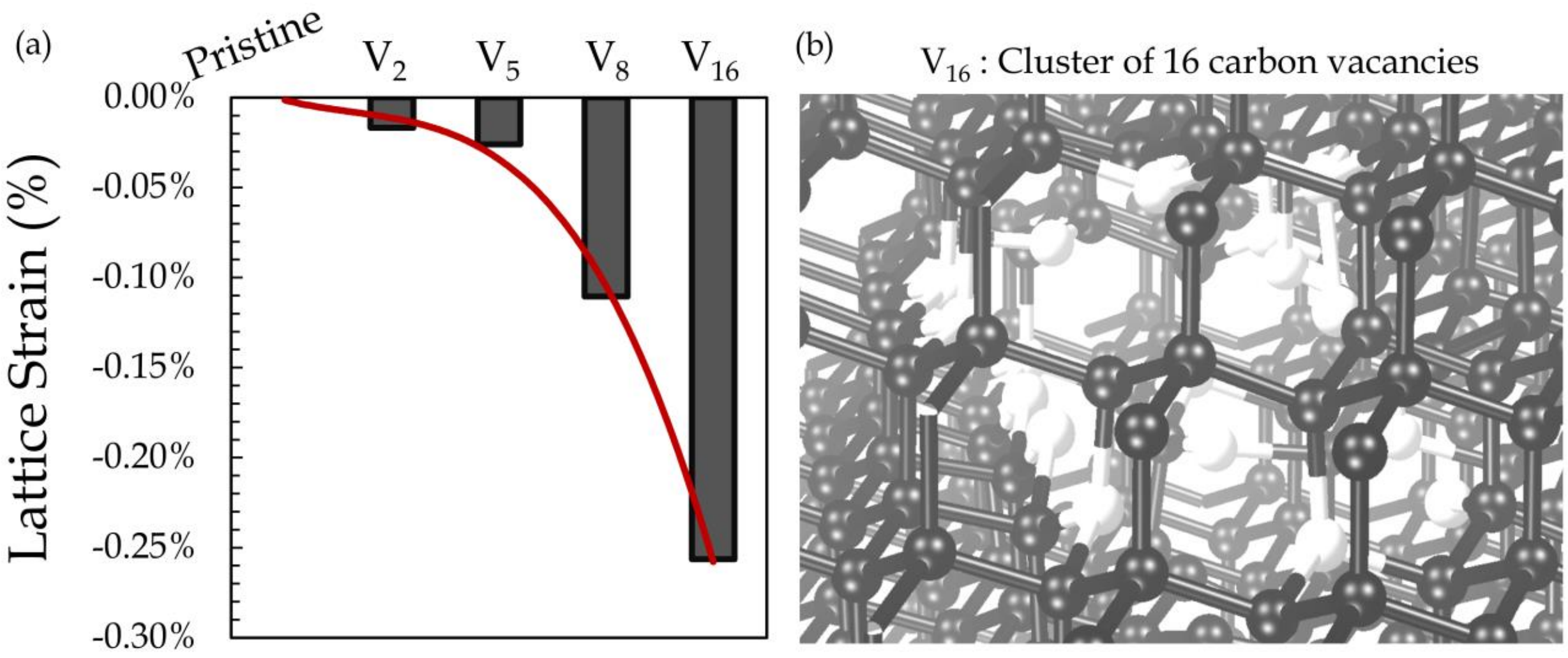

***Fig. S5.*** *(a)Vacancy clusters induce lattice compression in diamond, and increasing size of the cluster induces an even larger compressive strain. (b) The white colored atoms represents the carbon atoms in a vacancy cluster which show drastic shift of the Wyckoff coordinates and result in the shrinkage of the lattice parameters.*

The TEM observations revealed a reduction in the number of dislocations after annealing and a slight increase in the EELS onset of the energy loss counts (Fig. S6), which may be interpreted as an increasing bandgap. We also observe a reduction in birefringence, or strain due to annealing of the CVD diamonds, as seen under polarized light (Fig. S6c). The increase in bandgap and reduction in tensile strain (based on reduced 468 nm PL and reduction of brown coloration) are in direct agreement with the theoretically calculated strain induced changes in bandgap (Fig. S4b), where reduction in tensile strain coincides with increasing bandgap, underscoring the importance of EELS calculations that should take this effect into account, although ignored so far [A8].

Positron annihilation lifetime spectroscopy (PALS) has long been used to probe vacancy-type defects and has remained a crucial evidence for the vacancy cluster related nature of the brown coloration in CVD diamond, with increased positron annihilation lifetimes typically interpreted as signatures of open-volume clusters [A9, A10]. However, emerging evidence suggests that this interpretation may be incomplete: even modest lattice strain (~1%) or hydrogen incorporation can significantly reduce electron density, leading to elevated positron lifetimes without requiring actual vacancies.

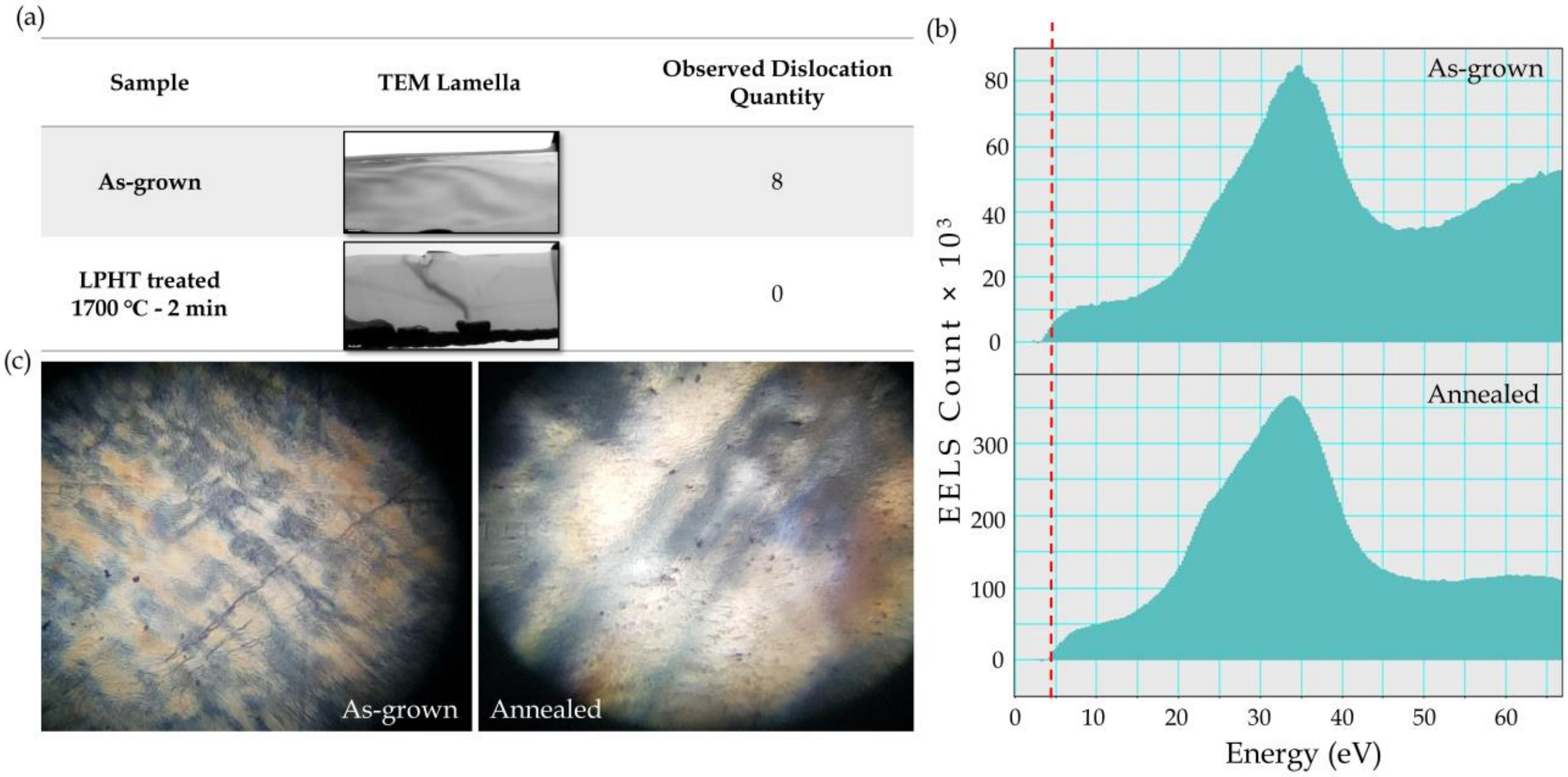

***Fig. S6.*** *(a) The TEM observations show a reduction in the dislocation due to annealing. (b) EELS data suggests a slight increase in bandgap due to annealing and (c) polarized microscopy reveals a drastic reduction of strain as a result of thermal treatment by LPHT.*

Elastic tensile strain lowers local electron density and deepens electrostatic potential wells, mimicking the lifetime effect of vacancies. Classic experiments on metals (e.g., Au, Al) and intermetallic compounds have shown that lattice expansion, regardless of the cause, tends to increase positron lifetime, as well [A11, A12]. Hydrogen plays a similar role through interstitial-induced lattice swelling. Kawasuso et al. [A13] showed that hydrogen uptake in Ti–Cr–V alloys increases both lattice parameters and positron lifetime, with the latter fully reversible upon hydrogen removal. Study of Zr–Nb alloys by Bordulev et al. [A14] reports significant increase in positron annihilation lifetime associated with hydrogen uptake, correlated with lattice dilation, despite the absence of confirmed vacancy clusters.

Our DFT simulations illustrate how a mere 1% strain perturbs the electron density in the diamond lattice (Fig. S7). The electrostatic potential well in a diamond slab deepens under 1% tensile strain (orange) compared to the unstrained lattice (blue), consistent with electrons being more delocalized (Fig. S7b-7c). Under 1% tension, low-electron-density regions (red isosurfaces at 0.14 e/Å$^3$) expand and connect, while high-density regions (blue isosurfaces at 1.60 e/Å$^3$) around atoms shrink – as if from an electron density perspective, extra "free volume" has emerged between atoms. Conversely, 1% compression squeezes these void regions (red) and enhances electron density overlap (blue), which directly result in variations of positron lifetime, which depends on electron density.

Altogether, this body of work urges a more nuanced analysis of positron lifetimes—particularly in strained or hydrogenated diamonds—where lattice expansion alone may be sufficient to explain observed lifetime shifts. Therefore, considering the findings of this study and the lowered electron density of diamond upon tensile strain (Fig. S7), the assignment of enhanced positron annihilation lifetime to vacancy clusters should be reconsidered in light of the possible contributions of H-induced lattice expansion.

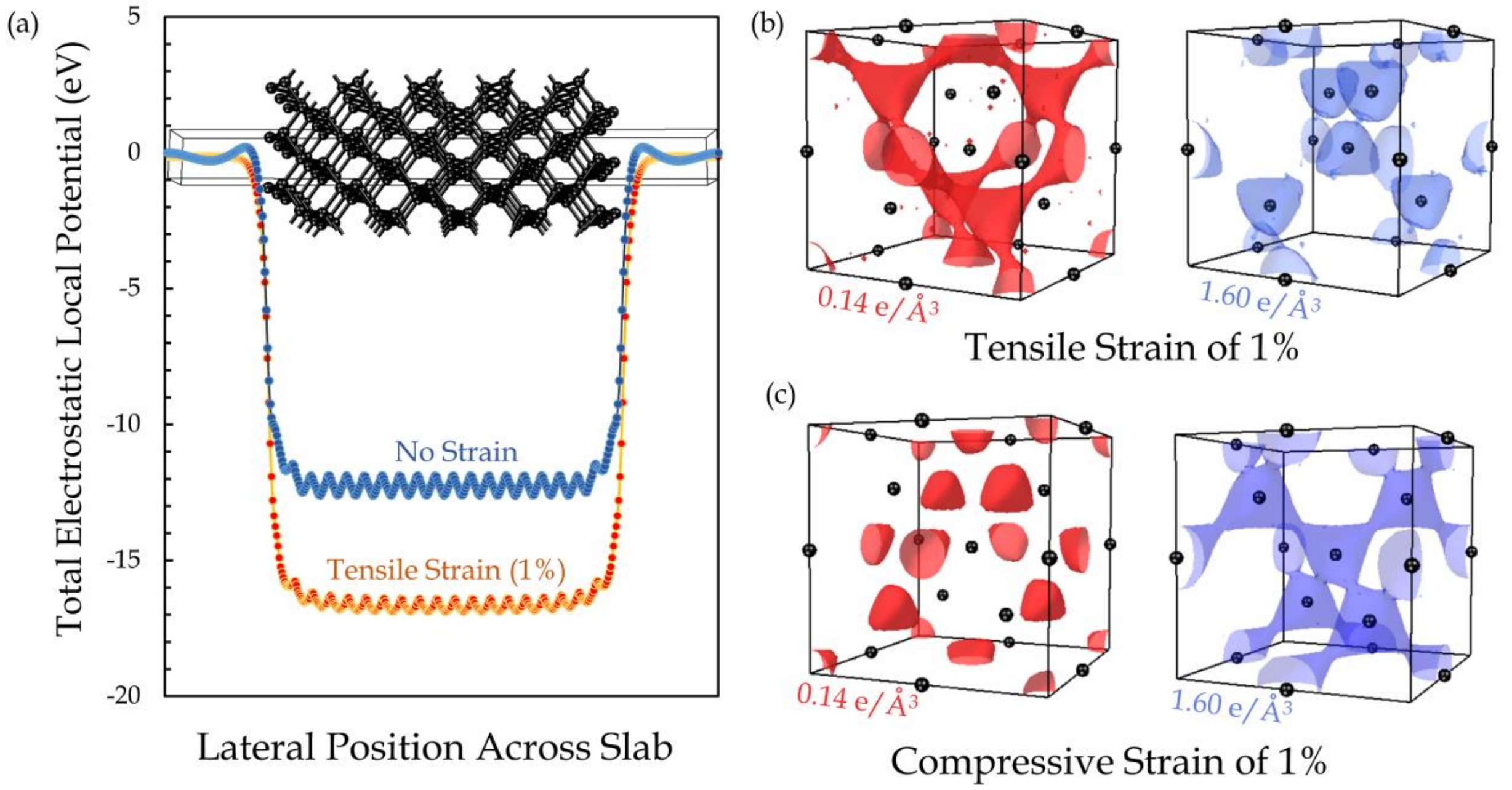

***Fig. S7.*** *(a) The electrostatic potential diamond slab deepens under 1% tensile strain (orange) compared to the unstrained lattice (blue). (b) Under 1% tension, low-electron-density regions (red isosurfaces at 0.14 e/Å³) expand and connect, while high-density regions (blue isosurfaces at 1.60 e/Å³) shrink, (c) Conversely, 1% compression squeezes these void regions (red) and enhances electron density overlap (blue), shortening the positron lifetime. This confirms that even elastic strain on the order of 1% can significantly alter electron densities in the lattice, directly affecting positron annihilation characteristics.*

## 3. Benchmarking of calculation methodology for PL lineshape

The theoretical calculation of photoluminescence lineshape is based on the methodology of Alkauskas et al. [A15], as implemented by the PyPhotonics code [A16] following calculation of the dynamical matrix for the ground and excited states of a defect. To verify the accuracy and reliability of the approach used, the PL lineshape and the ZPL for the $NV^-$ was initially calculated, and compared with the prior theoretically calculated PL and experimental data [A15]. The methodology used in our study has captured the PL lineshape of the $NV^-$ and the accuracy is acceptable (Fig. S8a).

Apart from the suggested assignment of the 468 nm center to NVH defect, there is an internal transition of $NVH^-$ in the IR region which is described here (Fig. S8b) for future experimental identification. The neutral NVH exhibits S=1 triplet 3A'' many body electronic state with half occupied 1a' and 1a'' levels in spin down channel. Trapping of an additional electron on the 1a'' level leads to the $NVH^{1-}$, which is a spin doublet. This charge state exhibits possible spin conserving internal transition 2A'' → 2A' in the spin minority channel that falls within IR range.

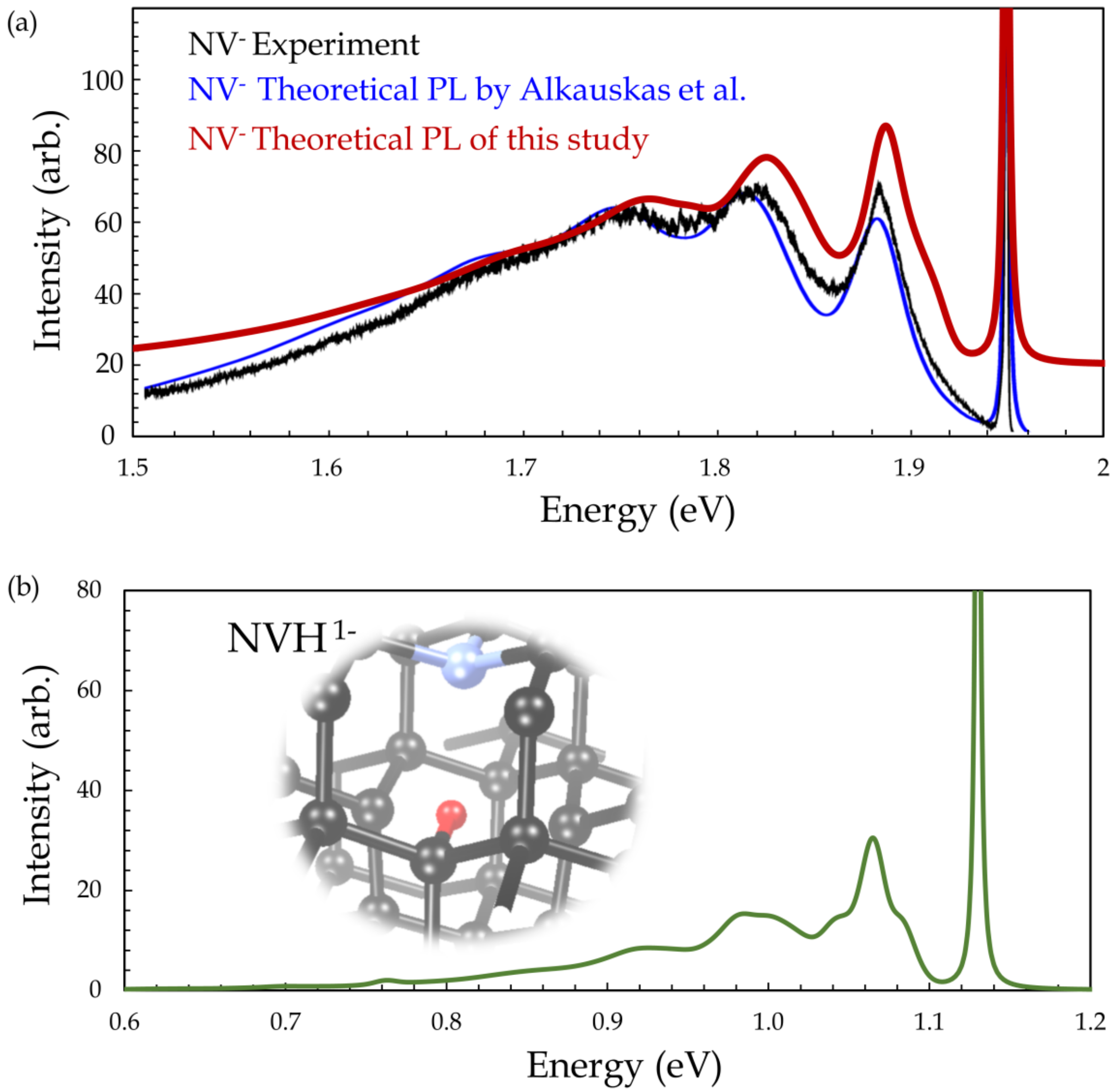

***Fig. S8.*** *(a) The calculated NV⁻PL lineshape by Alkasukas et al[A15] and the experimental data [A15] is compared to the PL lineshape we have calculated to benchmark the reliability of the methods used in this study. (b) We also report an internal radiative transition of NVH⁻ with a ZPL in the IR region and the phonon side band shown here.*

## 4. $N_xVH_y$ point defect induced coloration and the color transformations during annealing

The dielectric function for each defect and the dominant charge states have been calculated through the HSE functional based on methodology of Gajdoš et al. [A17], which are then used for the estimation of absorption spectra for each defect. The attenuation coefficients as a function of wavelength for the defects in various charge states considered in this study are shown in Fig. S9.

We find that the 270 nm band is not just a result of the $N^-$, as previously suggested, but rather $N^-$ is a key contributor to this band. $NVH^{2-}$, and $N_2VH^-$ are key defects that play a role in the absorption of the 270 nm band as well. The findings also confirm the NVH assignment of the 520 nm band that was suggested by Khan et al. [A18], however, the band is not a product of just one defect but rather a defect ensemble, in which the NVH plays a crucial role. Moreover, the spectra shown below show the possibility of generating a variety of brown to pinkish brown and pink shades through different mixtures of these defects, explaining not just the brown coloration of as-grown CVD diamond, but also the brown to brownish pink transformations that occur during LPHT treatments of brown diamond; dissociation of $NVH_x$ to $NVH^-$ and finally $NV^-$ that cause pinkish hues.

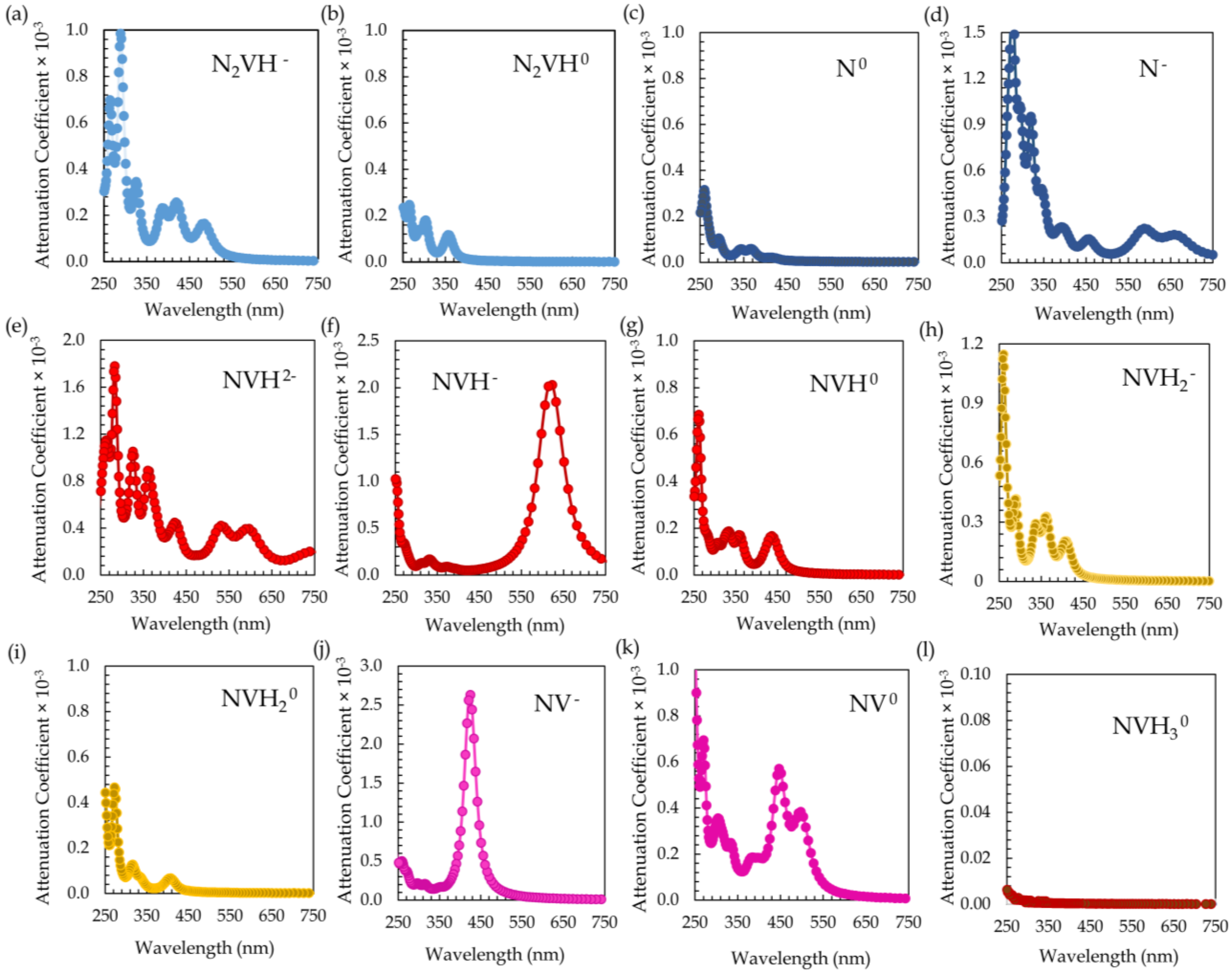

***Fig. S9***. *The UV-VIS absorption spectra of different defects calculated by hybrid DFT are shown here, for (a)* $N_2VH^-$ *(b)* $N_2VH^0$*, (c)* $N^0$*, (d)* $N^-$*, (e)* $NVH^{2-}$*, (f)* $NVH^-$*, (g)* $NVH^0$*, (h)* $NVH_2^-$*, (i)* $NVH_2^0$*, (j)* $NV^-$*, (k)* $NV^0$*, (l)* $NVH_3$

**Supplementary References**